\documentclass[pmlr,twoside,tablecaptionbottom]{pgm-jmlr}
\usepackage{pgm-pmlr-dat}
\firstpageno{1}

\usepackage[utf8]{inputenc}
\inputencoding{utf8}

\usepackage{mathtools} 
\usepackage{booktabs} 
\usepackage{tikz} 

\usepackage{subcaption}

\definecolor{mypink1}{rgb}{0.858, 0.188, 0.478}
\definecolor{mypink2}{RGB}{219, 48, 122}
\definecolor{mypink3}{cmyk}{0, 0.7808, 0.4429, 0.1412}
\definecolor{mygray}{gray}{0.6}
\definecolor{darkgreen}{HTML}{006622}

\definecolor{dagpink}{HTML}{FF7777}
\definecolor{dagblue}{HTML}{00A2E0}
\definecolor{daggreen}{HTML}{BED403}

\usepackage{fancyvrb}

\RecustomVerbatimCommand{\VerbatimInput}{VerbatimInput}%
{fontsize=\footnotesize,
 frame=lines,  
 framesep=2em, 
 rulecolor=\color{Gray},
 label=\fbox{\color{Black}IMF\_CP\_d32.dot},
 labelposition=topline,
 commandchars=\|\(\), 
 commentchar=*        
}

\usepackage[retainorgcmds]{IEEEtrantools}
\usepackage{paralist}

\title[$\rho$-GNF for Sensitivity Analysis to Unobserved Confounding]
{$\rho$-GNF: A Copula-based Sensitivity Analysis to \\Unobserved Confounding Using Normalizing Flows}

 \author{
 \Name{Sourabh Balgi} \Email{sourabh.balgi@liu.se}\\
 \addr STIMA, IDA, Linköping University, Sweden
 \AND
 \Name{Jose M. Pe\~na} \Email{jose.m.pena@liu.se}\\
 \addr STIMA, IDA, Linköping University, Sweden
  \AND
 \Name{Adel Daoud} \Email{adel.daoud@liu.se}\\
 \addr IAS, IEI, Linköping University, Sweden
 }

\begin{document}
\maketitle

\begin{abstract}
We propose a novel sensitivity analysis to unobserved confounding in observational studies using copulas and normalizing flows. Using the idea of interventional equivalence of structural causal models, we develop $\rho$-GNF ($\rho$-graphical normalizing flow), where $\rho{\in}[-1,+1]$ is a bounded sensitivity parameter. This parameter represents the back-door non-causal association due to unobserved confounding, and which is encoded with a Gaussian copula. In other words, the $\rho$-GNF enables scholars to estimate the average causal effect (ACE) as a function of $\rho$, while accounting for various assumed strengths of the unobserved confounding. The output of the $\rho$-GNF is what we denote as the $\rho_{curve}$ that provides the bounds for the ACE given an interval of assumed $\rho$ values. In particular, the $\rho_{curve}$ enables scholars to identify the confounding strength required to nullify the ACE, similar to other sensitivity analysis methods (e.g., the E-value). Leveraging on experiments from simulated and real-world data, we show the benefits of $\rho$-GNF. One benefit is that the $\rho$-GNF uses a Gaussian copula to encode the distribution of the unobserved causes, which is commonly used in many applied settings. This distributional assumption produces narrower ACE bounds compared to other popular sensitivity analysis methods. 
\end{abstract}

\section{Introduction}\label{sec:intro}
\looseness=-1
Epidemiologists, sociologists, economists, and other applied scientists, often leverage randomized controlled trials (RCTs), as  RCTs provide the safest methodological route to disentangle cause and effect. RCTs are the \emph{gold-standard} as they require the least assumptions~\citep{wright1921correlationcausation, fisher1936designofexperiments, cox1958planning, Imbens2015CIinSocialscience}.
Most importantly, by randomizing which experimental subjects (e.g., people, villages, schools) should take the treatment and which subjects should abstain, an RCT provides unconditional ignorability or exchangability or unconfoundedness.
\textit{Unconfoundedness} implies no unobserved confounders in the causal system of interest: no unobserved common causes of the treatment and the outcome. 
When unconfoundedness is satisfied, scholars can calculate the causal effect of interest from collected data; that means that the causal quantity is identified ~\citep{ROBINS1986gcom, Rubin1990unconfoundedness, robins2008ipw, pearl2009causality, hernan2009ipw}.
However, despite the importance of the RCT design, it remains infeasible for a slew of applied settings. It may be costly to implement (e.g., testing a population-wide medicine); it may be unethical (e.g., testing a new drug); or, it may be impractical to implement (e.g., testing a social policy across the world). Therefore, applied researchers often rely on observational data -- which are often secondary data sources with no treatment randomization and where the experimenter had no control over the data generating process. Yet when using observational data, scholars make themselves susceptible for failing to satisfy the unconfoundedness assumption, even when some confounders are observed.

\looseness=-1
Because the unconfoundedness assumption is so critical and at the same time untestable in observational studies~\citep{Rubin1990unconfoundedness, d2020accounting}, methodologists (statisticians, computer scientist, and others) have developed various frameworks for stress testing how causal effect estimates change under varying strength of unconfoundedness failure. These sort of tests are named \emph{sensitivity analysis} ~\citep{schlesselman1978assessing, manski1990sensitivitybounds, imbens2003sensitivity, brumback2004sensitivity, vanderweele2011bias}, also known as \emph{bias analysis} in epidemiology~\citep{cornfield1959smoking, cochran1973controlling, rothman2008modern, lash2009applying}.
Nonetheless, existing sensitivity analysis frameworks are limited in at least three ways, and our proposed $\rho$-GNF improve on these limitations, thereby moving the state-of-the-art forward. 

First, with the idea of interventional equivalence of structural causal models (SCMs), we propose a deep-learning method for sensitivity analysis based on the graphical normalizing flow (GNF)~\citep{wehenkel2020GNF}, causal graphical normalizing flow (c-GNF)~\citep{balgi2022cgnf} and Causal Normalizing Flow (CNF)~\citep{JavaloySV23}, because of GNF's attractive properties of non-linearity and invertibility for counterfactual inference and the similarities to the most well studied and popular elliptical copula, i.e., Gaussian copula. Hence, we aptly name the model $\rho$-GNF, where $\rho{ \in }[-1,+1]$ is the bounded sensitivity parameter of the Gaussian copula that controls the degree of unconfoundedness between the observed treatment and outcome. Unlike most sensitivity analysis methods where the sensitivity parameters are unbounded and difficult to specify and interpret, $\rho$ is bounded in the range $[-1,+1]$ and it represents the non-causal dependence between the treatment and outcome. 
Second, we show that with interventional equivalence, $\rho$-GNF enables us to estimate the interventional causal effects such as average causal effect (ACE) as a function of $\rho$. We call this the $\rho_{curve}$, and this curve enables us to identify the ACE bounds and analyze the confounding strength required to explain away the causal effect. Thus, the $\rho_{curve}$ enables us to provide bounds for the causal effect given a specific interval of $\rho{ \in }[-1,+1]$ that the domain expert considers appropriate. We also define $\rho_{value}$ as the value of $\rho$ that explains away the causal effect and demonstrate its similarities to the widely used E-value~\citep{vanderweele2017sensitivity}.
Third, we empirically demonstrate sharp and narrower bounds compared to the widely popular assumption-free bounds through simulated as well as real-world experiments. Unlike existing sensitivity analysis methods that provide bounds either only for discrete or continuous outcomes, $\rho$-GNF accommodates both discrete and continuous outcomes.

\looseness=-1
Our work proceeds as follows. After briefly reviewing related literature in Section~\ref{sec:background}, we define our notation and the main objective of ACE estimation, using the idea of interventional equivalence of SCMs with bivariate Gaussian copula in Section~\ref{sec:notation_problem_definition_construction}. We perform the identification and estimation of the ACE and analyse the sensitivity to the different degree of unconfoundedness in Section~\ref{sec:ACE}-\ref{sec:evalue_rho_value}, via the sensitivity parameter $\rho$. In Section~\ref{sec:experiments}, we present our results with simulated as well as real-world data under different settings of outcome variable (i.e., continuous or binary or categorical), and compare them with the popular assumption-free (AF) bounds.
Finally, in Section~\ref{sec:conslusion}, we conclude with discussing the key contributions of our $\rho$-GNF method in encouraging the use of sensitivity analysis when working with non-randomized observational data.

\subsection{Background and Related Work}\label{sec:background}
\looseness=-1
The sensitivity analysis literature can be roughly categorized into two streams:
(i) identify the \emph{bounds} of the causal effect as functions of some sensitivity parameters that encode the strength of the unobserved confounders~\citep{robins1989analysis, manski1990sensitivitybounds, vanderweele2011bias, ding2016sensitivity, sjolander2020note, sjolander2021novel, pena2022simple}; and (ii) identify how large the influence of the unobserved confounders needs to be to \emph{explain-away} the causal effect~\citep{imbens2003sensitivity, vanderweele2017sensitivity, veitch2020sense, sjolander2022values}.
While ~\citet{robins1989analysis, manski1990sensitivitybounds} provide assumption-free (AF) bounds of the causal effect for binary outcome, more recent methods such as~\citet{ilse2021efficient} extend the bounds to categorical outcomes.
Other methods provide bounds as functions of sensitivity parameters to be tuned by a domain expert~\citep{sjolander2020note, sjolander2021novel, pena2022simple}.
~\citet{cinelli2019sensitivity, cinelli2020making} study sensitivity to unobserved confounding in a linear SCM setting.
In contrast to the \emph{bounds} stream, there are methods that fall under the \emph{explain-away} stream.
For example,~\citet{imbens2003sensitivity, vanderweele2017sensitivity} reason on the lines of the minimum strength of the unmeasured confounder that is needed, conditional on the measured confounders, to explain-away the estimated causal effect.
Similar to~\citet{imbens2003sensitivity} and the E-value~\citep{vanderweele2017sensitivity}, the recently developed \emph{Austen plots}~\citep{veitch2020sense} identify the influence of the confounding needed to explain a specific amount of bias in the causal effect estimate. 

\looseness=-1
While there exists a wide spectrum of sensitivity analysis methods with unique advantages, they are not without limiting assumptions. For example,~\citet{robins1989analysis, manski1990sensitivitybounds, sjolander2020note, sjolander2021novel, pena2022simple} require the outcome variable being binary, and other methods assume a specific type of parametric model, e.g.,~\citet{cinelli2019sensitivity, cinelli2020making} assume a linear parametric model. While some methods offer sharp ACE bounds~\citep{robins1989analysis, manski1990sensitivitybounds, sjolander2020note, sjolander2021novel, pena2022simple}, method such as~\citet{vanderweele2017sensitivity} may result in wider bounds than the AF bounds, as shown by~\citet{ioannidis2019limitations} and~\citet{sjolander2020note}. Some methods are exclusively suited for specific causal estimands, e.g., ACE or conditional ACE (CACE) or mediation effects~\citep{tchetgen2012semiparametric, lindmark2018sensitivity}.
While~\citet{robins1989analysis, manski1990sensitivitybounds} offer no sensitivity parameters that can explain a certain causal effect,~\citet{vanderweele2017sensitivity, veitch2020sense} offer multiple parameters that are unbounded and hard to specify for the domain analyst~\citep{ioannidis2019limitations}.

To summarize, even though there exists several sensitivity analysis frameworks, there is still a lack of unifying method that is flexible enough that can suit many different types of observational data, with easy-to-use sensitivity parameters, and that can be applied to not only binary outcome variables but also categorical and continuous outcomes. 
Moreover, it's imperative to establish a method enabling researchers to specify distributional assumptions regarding the unobserved causes within the causal system under study. Such assumptions enable tighter ACE bounds, enhancing the certainty of outcomes. Our $\rho$-GNF method targets all these lacks. Our method use deep neural networks, allowing for maximum flexibility and non-linearity. $\rho$-GNF provides a single, bounded sensitivity parameter $\rho{ \in }[-1,+1]$ that is easily interpreted as the measure of non-causal association due to unobserved confounders. 
Thus, our method enhances an applied researcher's causal toolbox, and shows how deep learning can further causal inference.

\section{Notation and Problem Definition}\label{sec:problemdefinition,assumptions,notations}
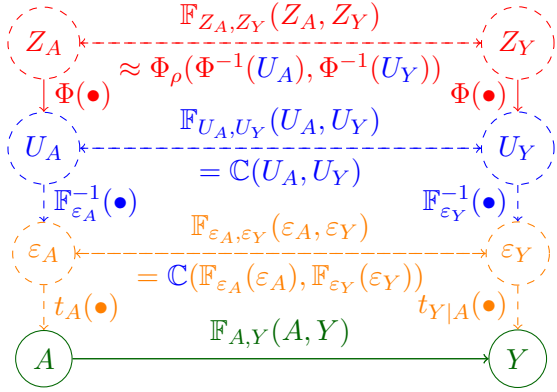
\begin{figure}[t!]
\begin{center}
\begin{tikzpicture}[scale = 1.4] 
\tikzset{vertex/.style = {shape=circle,draw,minimum size=1.5em}}
\tikzset{edge/.style = {->}}
\node[red,vertex,dashed] (Z_A) at (0,2) {$Z_A$};
\node[red,vertex,dashed] (Z_Y) at (4.5,2) {$Z_Y$};
\node[blue,vertex,dashed] (U_A) at (0,1) {$U_A$};
\node[blue,vertex,dashed] (U_Y) at (4.5,1) {$U_Y$};
\node[orange,vertex,dashed] (e_A) at (0,0) {$\varepsilon_A$};
\node[orange,vertex,dashed] (e_Y) at (4.5,0) {$\varepsilon_Y$};
\node[black!60!green,vertex] (A) at (0,-1) {$A$};
\node[black!60!green,vertex] (Y) at (4.5,-1) {$Y$};
\draw[black!60!green,edge] (A) to node[above,sloped] {{$\mathbb{F}_{A,Y}(A,Y)$}} (Y);
\draw[orange,edge,dashed] (e_A) to node[right] {$t_A(\bullet)$} (A);%
\draw[orange,edge,dashed] (e_Y) to node[left] {$t_{Y|A}(\bullet)$} (Y);
\draw[orange,edge,dashed] (e_Y) to node[above,sloped] {$\textcolor{orange}{\mathbb{F}_{\varepsilon_A,\varepsilon_Y}(\varepsilon_A,\varepsilon_Y)}$} (e_A);
\draw[orange,edge,dashed] (e_A) to node[below] {$=\textcolor{blue}{\textcolor{blue}{\mathbf{\mathbb{C}}}}(\textcolor{orange}{\mathbb{F}_{\varepsilon_A}(\varepsilon_A),\mathbb{F}_{\varepsilon_Y}(\varepsilon_Y)})$} (e_Y);
\draw[blue,edge,dashed] (U_A) to node[right] {$\mathbb{F}^{-1}_{\varepsilon_A}(\bullet)$} (e_A);
\draw[blue,edge,dashed] (U_Y) to node[left] {$\mathbb{F}^{-1}_{\varepsilon_Y}(\bullet)$} (e_Y);
\draw[blue,edge,dashed] (U_Y) to node[above,sloped] {$\mathbb{F}_{U_A,U_Y}(U_A,U_Y)$} (U_A);
\draw[blue,edge,dashed] (U_A) to node[below,sloped] {{$=\textcolor{blue}{\mathbf{\mathbb{C}}}(U_A,U_Y)$}} (U_Y);
\draw[red,edge] (Z_A) to node[right] {$\Phi(\bullet)$} (U_A);
\draw[red,edge] (Z_Y) to node[left] {$\Phi(\bullet)$} (U_Y);
\draw[red,edge,dashed] (Z_Y) to node[above] {$\textcolor{red}{\mathbb{F}_{Z_A,Z_Y}(Z_A,Z_Y)}$} (Z_A);
\draw[red,edge,dashed] (Z_A) to node[below] {{$\approx\textcolor{red}{\Phi_\rho(\Phi^{-1}(}\textcolor{blue}{U_A}\textcolor{red}{),\Phi^{-1}(}\textcolor{blue}{U_Y}\textcolor{red}{))}$}} (Z_Y);
\draw[black!60!green,edge] (A) to (Y);
\end{tikzpicture} 
\end{center}
\caption{
Observationally and interventionally equivalent SCM representations from Eqs.~\eqref{seq:eps dist scm1}-\eqref{seq:eps dist copula_confounding2}.
The solid-lined nodes indicate observed treatment/outcome variables and the dash-lined nodes indicate unobserved cause/noise variables. 
The \textcolor{darkgreen}{green}, \textcolor{orange}{orange}, \textcolor{blue}{blue} and \textcolor{red}{red} lines respectively indicate the desired front-door causal association \textcolor{darkgreen}{$A{\rightarrow}Y$}, non-causal association due to ``effective'' noises \textcolor{orange}{$\varepsilon_X$}, non-causal association due to uniform marginal noises \textcolor{blue}{$U_X$}$\in[0,1]$, non-causal association due to Gaussian marginal noises \textcolor{red}{$Z_X$}.
}
\label{sfig:copula_confounding_cgnf_rho}
\end{figure}

\looseness=-1
\noindent
Let us consider the standard \emph{canonical representation} (in Chapter 3.4 of~\citet{peters2017eci}) for the structural causal model (SCM)~\citep{wright1921correlationcausation, Haavelmo1943SEM, Goldberger1972SEMecon, Fienberg1975Introduction2SEM, TabarJYCFL2022ML4PP}, where $A$ is the treatment (or cause) and $Y$ is the outcome (or effect), and $\varepsilon_A$ and $\varepsilon_Y$ are their respective unobserved causes such that

\vspace{-5mm}\begin{IEEEeqnarray}{rCl}
A := t_A(\varepsilon_A)  \enspace, \enspace Y := t_Y(A,\varepsilon_Y) = t_{Y|A}(\varepsilon_Y) \enspace, \enspace (\varepsilon_A, \varepsilon_Y){\sim}\textcolor{orange}{\mathbb{F}_{\varepsilon_A,\varepsilon_Y}(\varepsilon_A,\varepsilon_Y)}&\enspace,\enspace\IEEEyesnumber\label{seq:eps dist scm1}
\end{IEEEeqnarray}
where $(\varepsilon_A, \varepsilon_Y)$ 
follows the joint CDF $\textcolor{orange}{\mathbb{F}_{\varepsilon_A,\varepsilon_Y}(\varepsilon_A,\varepsilon_Y)}$ and is not limited to the usual assumptions of standard normal or uniform random variables as in most common SCM definitions. 
Using the universality of the uniform (also know as the probability integral transform)~\citep{angus1994probability}, the noise variables $\varepsilon_A$ and $\varepsilon_Y$ of the SCM in Eq.~\eqref{seq:eps dist scm1} can equivalently be written in terms of uniform variables $U_A$ and $U_Y$ in the interval $[0,1]$ resulting in the SCM for Figure~\ref{sfig:copula_confounding_cgnf_rho} (\textcolor{orange}{orange}+\textcolor{blue}{blue}) as

\vspace{-5mm}\begin{IEEEeqnarray}{rCl}
A := t_A(\mathbb{F}_{\varepsilon_A}^{-1}(U_A))  \enspace, \enspace Y := t_{Y|A}(\mathbb{F}_{\varepsilon_Y}^{-1}(U_Y)) \enspace , \enspace (U_A, U_Y){\sim}\textcolor{blue}{\mathbb{F}_{U_A,U_Y}(U_A,U_Y)}&\enspace , \IEEEyesnumber\label{seq:eps dist copula_confounding}
\end{IEEEeqnarray}
where $\mathbb{F}_{\varepsilon_A}$ and $\mathbb{F}_{\varepsilon_Y}$ respectively denote the marginal CDFs of $\varepsilon_A$ and $\varepsilon_Y$, and $(U_A, U_Y)$ follows the joint CDF $\textcolor{blue}{\mathbb{F}_{U_A,U_Y}(U_A,U_Y)}$ with uniform marginals in [0,1].
From the universality of the uniform, Eq.~\eqref{seq:eps dist copula_confounding} can further be simplified in terms of $\mathbb{F}_{A}$ and $\mathbb{F}_{Y|A}$ that denote the marginal CDFs of $A$ and $Y$ conditioned on $A$, respectively as below.

\vspace{-5mm}\begin{IEEEeqnarray}{rCl}
A := \mathbb{F}_{A}^{-1}(U_A)  \enspace, \enspace Y := \mathbb{F}_{Y|A}^{-1}(U_Y) \enspace , \enspace  (U_A, U_Y) {\sim}\textcolor{blue}{\mathbb{F}_{U_A,U_Y}(U_A,U_Y)}&\enspace . \IEEEyesnumber\label{seq:eps dist copula_confounding1}
\end{IEEEeqnarray}
\looseness=-1
Further, we represent $U_A$ and $U_Y$ in Eq.~\eqref{seq:eps dist copula_confounding1} as transformation of standard normal variables using the CDF of the standard normal $\Phi$, as shown in Figure~\ref{sfig:copula_confounding_cgnf_rho} (\textcolor{orange}{orange}+\textcolor{blue}{blue}+\textcolor{red}{red}), as 

\vspace{-5mm}\begin{IEEEeqnarray}{rCl}
A := \mathbb{F}_{A}^{-1}(\Phi(Z_A)) \enspace, \enspace Y := \mathbb{F}_{Y|A}^{-1}(\Phi(Z_Y)) \enspace,\enspace  (Z_A, Z_Y){\sim}\textcolor{red}{\mathbb{F}_{Z_A,Z_Y}(Z_A,Z_Y)}&\enspace ,\IEEEyesnumber\label{seq:eps dist copula_confounding2}
\end{IEEEeqnarray}
where $(Z_A, Z_Y)$ follows the joint CDF $\textcolor{red}{\mathbb{F}_{Z_A,Z_Y}(Z_A,Z_Y)}$ with standard normal marginals. 
Eqs.~\eqref{seq:eps dist scm1}-\eqref{seq:eps dist copula_confounding2} represent observationally and interventionally equivalent SCMs from Figure~\ref{sfig:copula_confounding_cgnf_rho}, but with different unobserved noises and corresponding joint CDFs. 
\textit{Observational equivalence} means that the researcher specified model yields the same distribution as observed by the true (nature or God) data-generating process {$\mathbb{F}_{A,Y}(A,Y)$}; \textit{interventional equivalence} means that the researcher specified model follows the same interventional distribution {$\mathbb{F}_{Y}(Y|do(a))$} and {$\mathbb{F}_{A}(A|do(y))$}~\citep{mooij2016distinguishing}. 
Since these noises (distributions) are unknown or unobserved, the completeness of $do$-calculus~\citep{tian2002unconfoundedchildrencriteria, HuangV06docalculus, pearl2012docalculus} states that the causal effects of interest are not identifiable from any of the equivalent SCMs in Eqs.~\eqref{seq:eps dist scm1}-\eqref{seq:eps dist copula_confounding2}, without further assumptions. 
To achieve causal effect/estimand identification and bounds, in the subsequent sections, we propose to use a bivariate Gaussian copula, to model the back-door non-causal noise distributions so that the causal effects can be parametrically estimated using deep-neural-network-inspired normalizing flows trained only on observational data.

\subsection{Representing the Back-Door Non-Causal Association using a Bivariate Gaussian Copula}\label{sec:notation_problem_definition_construction}
A copula is a multivariate distribution function defined on the unit hypercube with uniform marginals~\citep{sklar1959fonctions, sklar1973random}. 
As the name suggests, a copula \emph{`ties'} or \emph{`links'} or \emph{`couples'} a multidimensional joint distribution to its marginals~\citep{nelsen2007introduction}.
From the result of Sklar's Theorem, we have that the bivariate joint CDF $\textcolor{blue}{\mathbb{F}_{U_A,U_Y}(U_A,U_Y)}$ in Eqs.~\eqref{seq:eps dist copula_confounding} and~\eqref{seq:eps dist copula_confounding1} with uniform marginals in [0,1] can be represented using a bivariate copula $\textcolor{blue}{\textcolor{blue}{\mathbf{\mathbb{C}}}(U_A,U_Y)}$.

\vspace{-5mm}\begin{IEEEeqnarray}{rCl}
\textcolor{blue}{\mathbb{F}_{U_A,U_Y}(U_A,U_Y)} &=& \textcolor{blue}{\textcolor{blue}{\mathbf{\mathbb{C}}}(U_A,U_Y)}\IEEEyesnumber\IEEEyessubnumber\enspace.\label{seq:sklarstheorem}\\
\textcolor{orange}{\mathbb{F}_{\varepsilon_A,\varepsilon_Y}(\varepsilon_A,\varepsilon_Y)} &=& \textcolor{blue}{\textcolor{blue}{\mathbf{\mathbb{C}}}}(\textcolor{orange}{\mathbb{F}_{\varepsilon_A}(\varepsilon_A),\mathbb{F}_{\varepsilon_Y}(\varepsilon_Y)})\IEEEyessubnumber\enspace.\label{seq:sklarstheorem1}\\
\textcolor{red}{\mathbb{F}_{Z_A,Z_Y}(Z_A,Z_Y)} &=& \textcolor{blue}{\textcolor{blue}{\mathbf{\mathbb{C}}}}(\textcolor{red}{\Phi(Z_A),\Phi(Z_Y)})\IEEEyessubnumber\enspace.\label{seq:sklarstheorem2}
\end{IEEEeqnarray}
Eqs.~\eqref{seq:sklarstheorem1} and~\eqref{seq:sklarstheorem2} follow from the scale-invariance property of copula $\textcolor{blue}{\textcolor{blue}{\mathbf{\mathbb{C}}}(U_A,U_Y)}$ to strictly increasing transformations/CDFs $\textcolor{orange}{\mathbb{F}_{\varepsilon_A}}$, $\textcolor{orange}{\mathbb{F}_{\varepsilon_Y}}$, and $\textcolor{red}{\Phi}$.
\looseness=-1
The unknown copula $\textcolor{blue}{\mathbf{\mathbb{C}}(U_A,U_Y)}$ in Figure~\ref{sfig:copula_confounding_cgnf_rho} essentially models the non-causal back-door association, and the degree of the non-causal association between $U_A$ and $U_Y$ may be quantified using measures of association such as ~\citet{spearman1987proof, spearman2010proof}'s $\rho_S$ or~\citet{kendall1938new}'s $\tau_K$, where $\rho_S,\tau_K{\in}[-1,+1]$. 
From the scale-invariance property of $\rho_S$ and $\tau_K$ to the strictly increasing transformations $\textcolor{orange}{\mathbb{F}_{\varepsilon_A}}$, $\textcolor{orange}{\mathbb{F}_{\varepsilon_Y}}$, and $\textcolor{red}{\Phi}$, this measure of association between $U_A$ and $U_Y$ is the same between $\varepsilon_A$ and $\varepsilon_Y$, and between $Z_A$ and $Z_Y$, i.e., $\textcolor{orange}{\rho_S(\varepsilon_A, \varepsilon_Y}){=}\textcolor{blue}{\rho_S(U_A, U_Y)}{=}\textcolor{red}{\rho_S(Z_A,Z_Y)}{=}\textcolor{blue}{\rho_{S_{\textcolor{blue}{\mathbf{\mathbb{C}}}}}}$. 
In other words, the \textcolor{orange}{orange}, \textcolor{blue}{blue} and \textcolor{red}{red} back-door paths in Figure~\ref{sfig:copula_confounding_cgnf_rho} induce the same measure of non-causal association $\textcolor{blue}{\rho_{S_{\textcolor{blue}{\mathbf{\mathbb{C}}}}}}$ that can be represented using the copula $\textcolor{blue}{\textcolor{blue}{\mathbf{\mathbb{C}}}(U_A,U_Y)}$.
This result intuitively follows as the SCMs are observationally and interventionally equivalent, i.e., same measures of total observed and causal associations are expected, implying the non-causal associations also to be the same.
As discussed above, for identifying the causal effect of interest, it is sufficient if we observe/known/hypothesize the noises or the copula to adjust for the non-causal back-door path. 
However, the copula $\textcolor{blue}{\mathbf{\mathbb{C}}(U_A,U_Y)}$, although uniquely exists because of the strictly increasing continuous transformations $\textcolor{orange}{\mathbb{F}_{\varepsilon_A}}$, $\textcolor{orange}{\mathbb{F}_{\varepsilon_Y}}$, and $\textcolor{red}{\Phi}$, remains unknown and cannot be estimated from observational data as the noises are unobserved.
Since the copula $\textcolor{blue}{\mathbf{\mathbb{C}}(U_A,U_Y)}$ is unknown and unlearnable, it is inevitable to make assumptions about it to achieve causal effect identification. Specifically, $\textcolor{blue}{\mathbf{\mathbb{C}}(U_A,U_Y)}$ may be chosen from any of the vast families of copulas such as Archimedean copulas~\citep{ling2020deep} (Clayton, Frank, Gumbel, etc.), elliptical, or empirical copulas~\citep{nelsen2007introduction, salvadori2007extremes, durante2016principles, benali2021mtcopula}. 
Recent works such as~\citet{zheng2021copula, zheng2022sensitivity} have proposed sensitivity analysis with one of the most well studied and used elliptical copula, namely the Gaussian copula, but without the use of normalizing flows.
In our current work, we present our analysis by assuming and approximating the unknown copula $\textcolor{blue}{\mathbf{\mathbb{C}}(U_A,U_Y)}$ with the Gaussian copula, while extending the normalizing flows with monotonic transformers that are recently shown to be universal non-linear SCM approximators~\citep{Huang2018NAF, wehenkel2019UMNN, wehenkel2020GNF, balgi2022cgnf}. 

\looseness=-1
The Gaussian copula has been widely used in the fields of quantitative finance~\citep{cherubini2004copula, salmon2009recipe, mackenzie2014formula}, hydrology research~\citep{renard2007use, zhang2019copulas}, logistics~\citep{kumar2019copula}, astronomy~\citep{takeuchi2010constructing}, and similar fields~\citep{nelsen2007introduction, salvadori2007extremes, durante2016principles}. 
This is one of the several motivations for the particular selection of the Gaussian copula.
The assumption and approximation of the unknown copula $\textcolor{blue}{\mathbf{\mathbb{C}}(U_A,U_Y)}$ in Eqs.~\eqref{seq:sklarstheorem}-\eqref{seq:sklarstheorem2} with a Gaussian copula $\textcolor{red}{\Phi_\rho(\Phi^{-1}(}\textcolor{blue}{U_A}\textcolor{red}{),\Phi^{-1}(}\textcolor{blue}{U_Y}\textcolor{red}{))}$ achieves causal effect identification as the back-door non-causal association between $Z_A$ and $Z_Y$ may be identified as \\
$\textcolor{red}{\mathbb{F}_{Z_A,Z_Y}(Z_A,Z_Y)}{=}\textcolor{blue}{\textcolor{blue}{\mathbf{\mathbb{C}}}}(\textcolor{red}{\Phi(Z_A),\Phi(Z_Y)}){\approx}\textcolor{red}{\Phi_\rho(Z_A,Z_Y)}$,
where $\textcolor{red}{\rho}{\in}[-1,+1]$ is 
the Pearson's correlation between $Z_A$ and $Z_Y$. 
As the copula and the noises $\varepsilon_X$ and $U_X$ in Figure~\ref{sfig:copula_confounding_cgnf_rho} may exhibit non-linear dependence, it is more appropriate to equivalently represent the linear Pearson's correlation parameter $\textcolor{red}{\rho}$ in the Gaussian copula $\textcolor{red}{\Phi_\rho}$ in terms of the non-linear measure of association $\textcolor{blue}{\rho_{S_{\mathbf{\mathbb{C}}}}}$ (or $\textcolor{blue}{\tau_{K_{\mathbf{\mathbb{C}}}}}$) using the following results for bivariate Gaussian copula~\citep{kruskal1958ordinal, meyer2013bivariate}.
This measure of back-door non-causal association $\textcolor{blue}{\rho_{S_{\mathbf{\mathbb{C}}}}}$ due to the Gaussian copula assumption 
equates to the Gaussian copula parameter $\textcolor{red}{\rho}$ as $\textcolor{red}{\rho}{=}2\sin(\pi\textcolor{blue}{\rho_{S_{\mathbf{\mathbb{C}}}}}/6)$ (i.e., $\textcolor{blue}{\rho_{S_{\mathbf{\mathbb{C}}}}}{\approx}\textcolor{red}{\rho}$).
With the Gaussian copula $\textcolor{red}{\Phi_\rho}$ assumption in Eq.~\eqref{seq:sklarstheorem}-\eqref{seq:sklarstheorem2}, the back-door non-causal association in Eqs.~\eqref{seq:eps dist scm1}-\eqref{seq:eps dist copula_confounding2} is known, 
signifying that the causal effects are now identifiable under the assumed copula. Moreover, they can be estimated from a given observational dataset $\{(A^\ell,Y^\ell)\}^{N_{train}}_{\ell{ = }1}$ to train a parametric model as the proposed $\rho$-GNF in Figure~\ref{sfig:copula_confounding_cgnf_rho} by rewriting Eq.~\eqref{seq:eps dist copula_confounding2} as

\vspace{-5mm}\begin{IEEEeqnarray}{rCl}
\rho\text{-GNF} :: \enspace& A := \mathbb{T}_A^{-1}(Z_A;\theta_A)  \enspace, \enspace Y := \mathbb{T}_{Y|A}^{-1}(Z_Y;\theta_Y) \enspace,\enspace (Z_A, Z_Y) \sim \textcolor{red}{\Phi_\rho(Z_A,Z_Y)} \enspace,&\IEEEyesnumber\label{seq:eps dist copula_confounding_rho_gnf}
\end{IEEEeqnarray}
where $\mathbb{T}_A(\bullet;\theta_A)$ and $\mathbb{T}_{Y|A}(\bullet;\theta_Y)$ represent monotonic transformations
parameterized by deep neural networks $\theta{=}(\theta_A,\theta_Y)$ using the integration-based unconstrained monotonic neural network (UMNN) transformer~\citep{wehenkel2019UMNN}. 
The conditioning of $Y$ on its parent variable $A$ in $\mathbb{T}_{Y|A}(\bullet;\theta_Y)$ is done with the graphical conditioner from graphical normalizing flow (GNF)~\citep{wehenkel2020GNF} using the directed acyclic graph (DAG) $A{\rightarrow}Y$ assumed in Figure~\ref{sfig:copula_confounding_cgnf_rho}.
We aptly refer to our model with the Gaussian copula assumption in Eq.~\eqref{seq:eps dist copula_confounding_rho_gnf} as $\rho$-GNF due to the similarities to GNF and c-GNF that propose normalizing flows for observational density estimation and causal inference, but not for sensitivity analysis (herein lies our novelty over GNF or c-GNF). 
As any normalizing flow, the UMNN transformers and graphical conditioners are trained by maximizing the $\log$-likelihood of the observational training dataset $\{(A^\ell,Y^\ell)\}^{N}_{\ell{ = }1}$~\citep{wehenkel2020GNF, balgi2022cgnf}, for a fixed $\rho$.

\looseness=-1
In principle, any copula maybe assumed in place of Gaussian copula in Eqs.~\eqref{seq:sklarstheorem}-\eqref{seq:sklarstheorem2}. The Gaussian copula assumption, similar to normalizing flows~\citep{tabak2010nf,tabak2013nf,rezende2015variationalNF, Papamakarios2017MAF, papamakarios2021NF_pmi, kobyzev2020NF}, facilitates efficient computation of the $\log$-likelihood of the observational training dataset. Thus, enabling \textbf{computationally efficient training} of $\rho$-GNF in Eq.~\eqref{seq:eps dist copula_confounding_rho_gnf}. 
The Gaussian copula assumption further enables sampling $(Z_A, Z_Y)$ from $\textcolor{red}{\mathbb{F}_{Z_A,Z_Y}(Z_A,Z_Y)}{\approx}\textcolor{red}{\Phi_\rho(Z_A,Z_Y)}$ efficiently for the estimation of Monte-Carlo expectation in Eq.~\eqref{seq:ace gcopula mscm4}, thus enabling \textbf{computationally efficient inference}.
Most importantly,
the Gaussian copula assumption provides a single bounded sensitivity parameter $\rho{\in}[-1,+1]$ for sensitivity analysis that can be used to control/model/block/adjust the back-door non-causal association under which the ACE is identifiable.
Thus, enabling \textbf{simple and efficient sensitivity analysis}.
Our subsequent experiments and results show that the Gaussian copula assumption works well empirically, as~\citet{ilse2021efficient} also observe for multiple unobserved confounders, the joint distribution of the observed variables becomes increasingly Gaussian due to the central limit theorem.

\subsection{Sensitivity Analysis and Estimation of the Causal Estimand, i.e., ACE}\label{sec:ACE}
\looseness=-1
The main objective is to estimate the average causal effect (ACE), which can be expressed as 

\vspace{-5mm}\begin{IEEEeqnarray}{lr}
ACE_{\rho} = \mathbf{E}[Y_1{-}Y_0]=\mathbf{E}[Y_1]{-}\mathbf{E}[Y_0] & \enspace,\IEEEyesnumber\label{seq:ACE}
\end{IEEEeqnarray}
where $Y_a$ denotes the potential outcome under the intervention $A{:=}a$.
In practise, the estimation of the ACE is done by Monte-Carlo expectation estimation by drawing the samples from the interventional distributions to approximate $\mathbf{E}[Y_1]$ and $\mathbf{E}[Y_0]$ as indicated below in Eqs.~\eqref{seq:Z_l gcopula mscm4 abduction}-\eqref{seq:ace gcopula mscm4}, after having trained the $\rho$-GNF for a specific measure of unobserved confounding $\rho$ on the given observational dataset. 
\emph{The First Law of Causal Inference}~\citep{pearl1999probabilitiesofcausation, pearl2009abductionactionprediction, pearl2009causality, pearl2018bookofwhy} provides three steps, i.e, abduction, action and prediction, to estimate $\mathbf{E}[Y_a]$ and thus the $ACE_{\rho}$ in Eq.~\eqref{seq:ACE}.
  
    \vspace{-5mm}
    \begin{IEEEeqnarray}{C}
Z_Y^\ell{=}\mathbb{T}_{Y|A^\ell}({Y}^\ell;\theta_Y) \enspace \forall \enspace   \ell \in \{1,\ldots,N\}  \enspace. \IEEEyessubnumber\label{seq:Z_l gcopula mscm4 abduction}\\
Y^\ell_a{=}\mathbb{T}^{-1}_{Y|a}(Z^\ell_Y;\theta_Y) \enspace  \forall \enspace   \ell \in \{1,\ldots,N\}  \enspace.\IEEEyessubnumber\label{seq:Y_a gcopula mscm4}\\
ACE_{\rho}=\mathbf{E}[Y_1]{-}\mathbf{E}[Y_0]{\approx}\frac{\sum^{N}_{\ell{=}1} Y^{\ell}_1}{N} - \frac{\sum^{N}_{\ell{=}1} Y^{\ell}_0}{N}.\IEEEyessubnumber\label{seq:ace gcopula mscm4}
\end{IEEEeqnarray}

\subsection{Similarities to E-value}\label{sec:evalue_rho_value}
\looseness=-1
\citet{vanderweele2017sensitivity} propose the E-value as the minimum strength of association, on the risk ratio scale, that an unmeasured confounder would need to have with both the treatment and the outcome to fully explain away a specific causal treatment–outcome association, conditional on the measured covariates. A large E-value implies that considerable unmeasured confounding would be needed to explain away an effect estimate. A small E-value implies that little unmeasured confounding would be needed to explain away an effect estimate. 
Similar to the E-value, we propose the $\rho_{value}$ which represents the Gaussian copula parameter value that explains away the causal association between the observed treatment $A$ and the observed outcome $Y$. 
In other words, setting the Gaussian copula parameter $\rho{=}\rho_{value}$ results in $ACE_{\rho}{=}0$, i.e., $\mathbf{E}[Y_1]{=}\mathbf{E}[Y_0]$ in Eqs.~\eqref{seq:ACE} and~\eqref{seq:ace gcopula mscm4}, i.e., the potential outcomes are independent of the treatments/interventions. This implies that the strictly increasing transformation $\mathbb{T}^{-1}_{Y|A}$ modeling $Y$ in Eq.~\eqref{seq:eps dist copula_confounding_rho_gnf} is independent of $A$, i.e., we have $Y{=}\mathbb{T}^{-1}_{Y}(Z_Y)$. 

From the scale-invariance property of Spearman's correlation to strictly increasing transformations $\mathbb{T}^{-1}_{A}$ and $\mathbb{T}^{-1}_{Y}$, we have $\textcolor{red}{\rho_S(Z_A,Z_Y)}{=}\textcolor{blue}{\rho_{S_\mathbf{\mathbb{C}}}}{=}\textcolor{darkgreen}{\rho_S(A,Y)}{=}\textcolor{darkgreen}{\rho_{S_{Obs}}}$, where $\textcolor{darkgreen}{\rho_{S_{Obs}}}$ represents the observed Spearman's correlation between the observational data $\textcolor{darkgreen}{A}$ and $\textcolor{darkgreen}{Y}$. 
For a given observational dataset $A$ and $Y$, if the total observed association $\textcolor{darkgreen}{\rho_{S_{Obs}}}$ between $A$ and $Y$ is completely due to the non-causal association $\textcolor{blue}{\rho_{S_\mathbf{\mathbb{C}}}}$, there can be no contribution from the causal association, i.e., $ACE_{\rho}{=}0 \Leftrightarrow \rho{=}\textcolor{darkgreen}{\rho_{value}}$ where we have $\textcolor{darkgreen}{\rho_{value}}{=}2sin\left( \pi \textcolor{darkgreen}{\rho_{S_{Obs}}}/6\right)$. 
In Section~\ref{sec:experiments simcont}, we also experimentally verify that $\rho{=}\textcolor{darkgreen}{\rho_{value}} \Leftrightarrow ACE_{\rho}{=}0$. Hence, the E-value and $\rho_{value}$ represent the same concept, rather interpreted in different scales. 
Similar to the E-value that can be computed directly from the observational dataset~\citep{mathur2018website}, the $\rho_{value}$ is also identified as a single, bounded intuitive parameter that can be computed directly from Spearman's rho $\textcolor{darkgreen}{\rho_{S_{Obs}}}$ between $A$ and $Y$, i.e., $\rho_{value}{=}2\sin(\pi \textcolor{darkgreen}{\rho_{S_{Obs}}}/6)$. This further shows the similarities of the E-value and $\rho_{value}$.
Since the total observed association between $A$ and $Y$ remains constant for a given observational dataset, the $ACE_{\rho}$ varies with $\rho \in [-1,+1]$. We aptly refer this sensitivity plot of $ACE_{\rho}$ to $\rho$ for a given observational dataset as the $\rho_{curve}$. 

\looseness=-1
The $\rho_{value}$ equips an analyst/domain-expert to determine the sign of the ACE (i.e., whether the treatment is harmful or beneficial), which is arguably the most important part of the ACE and one of the ultimate goals of causal inference. 
Specifically, suppose the domain expert hypothesizes a measure of confounding in the interval $[\rho_{min},\rho_{max}]$. 
Thus, the $\rho_{curve}$ enables us to bound the true ACE to the narrower interval $[ACE_{\rho_{max}},ACE_{\rho_{min}}]$, which may in turn help us identify the most important insight of the causal inference, i.e., the sign of the true ACE. 
In particular, if $\rho_{min}{>}\rho_{value}$ (or $\rho_{max}{<}\rho_{value}$) then we may conclude that the true ACE is negative (or positive), as observed from the $\rho_{curve}$ in Figures~\ref{sfig:Jose_HVR_SCMs_050000_rho_curves_b},~\ref{fig:ndgp_0000_9500_subset}, and~\ref{fig:IMFCP_sensitivity_analysis_full}.
~Furthermore, unlike E-value that only indicates the strength of the association due to unobserved confounding, $\rho_{value}$ indicates both strength and the sign of the association due to unobserved confounding, i.e., \emph{positive} or \emph{negative} association between $A$ and $Y$.

\section{Experimental Setup, Results, Analysis and Discussion}\label{sec:experiments}
\looseness=-1
Our $\rho$-GNF is implemented in PyTorch~\citep{paszke2017pytorch}\footnote{The $\rho$-GNF code is available at \url{https://github.com/sobalgi/rhoGNF}.} by adapting the baseline code of UMNN~\citep{wehenkel2019UMNN} and GNF~\citep{wehenkel2020GNF}\footnote{GNF: \url{https://github.com/AWehenkel/Graphical-Normalizing-Flows}, UMNN: \url{https://github.com/AWehenkel/UMNN}.}. 
As normalizing flows are developed for continuous variables, we use the Gaussian dequantization trick from c-GNF~\citep{balgi2022cgnf} to model discrete variables into $\rho$-GNF. 
For the experiments, we present two different settings:
(i) simulated dataset with continuous outcomes in Section~\ref{sec:experiments simcont}, and
(ii) simulated dataset with binary outcomes in Section~\ref{sec:experiments simbin}. 
We additionally present the experiments using a real-world dataset with categorical outcomes analysing the impact of the IMF (International Monetary Fund) program on the degree of child poverty in the Global-South region by denoting the outcome as the total degree of child poverty ranging in 0-7 classes.

In practise, the $\rho_{curve}$ for a given observational dataset is obtained by varying the values of $\rho {\in}[-1,+1]$, i.e., we select $\rho{=}\{-0.99, -0.8, -0.6, -0.4, -0.2, 0.0, +0.2, +0.4, +0.6,\\ +0.8,$ $+0.99\}$ to train $\rho$-GNFs. We estimate the respective $ACE_{\rho}$ from Eqs.~\eqref{seq:ACE}-\eqref{seq:ace gcopula mscm4} to plot the $\rho_{curve}$ as interpolation of the points ($\rho, ACE_{\rho}$) shown in Figures~\ref{fig:Jose_HVR_SCMs_050000_rho_curves},~\ref{fig:ndgp_0000_9500_subset} and~\ref{fig:IMFCP_sensitivity_analysis_full}.
Our empirical ACE bounds are obtained as the infimum and supremum of the $\rho_{curve}$, i.e., $\inf[ACE_{\rho}] \leq ACE_{true} \leq \sup[ACE_{\rho}]$. We identify the $\rho_{value}$ that explains away the causal association for the benefit of domain expert/analyst.


\subsection{Experiments with Continuous Outcomes}\label{sec:experiments simcont}
\begin{table*}[ht!]
  \centering
  \begin{tabular}{l|ccc|ccc|rcl}
    \toprule
    $SCM_{\alpha,\beta,\delta}$ & $\alpha$  & $\beta$ & $\delta$ & $\textcolor{darkgreen}{\rho_{P_{Obs}}}$ & $\rho_{true}$  & $ACE_{true}$ &  $\rho$ &  $ \Leftrightarrow$ &  $ACE_\rho$ \\
    \midrule
$SCM_{1}$ &  0.2 & -0.6 & 0.72 & -0.55 & -0.71  & 0.2 & -0.71 &  $\Leftrightarrow$ &  0.2\\
$SCM_{2}$ &  0.0 & -0.4 & 0.52 & -0.55 & -0.55 & 0.0 & -0.55 &  $\Leftrightarrow$ &  0.0 \\
$SCM_{3}$ &  -0.2 & -0.2 & 0.40 & -0.55 & -0.32  & -0.2 & -0.32 &  $\Leftrightarrow$ &  -0.2  \\
\midrule
$SCM_{4}$ &  0.2 & 0.2 & 0.40 & 0.55 & 0.32  & 0.2  & 0.32 & $\Leftrightarrow$ & 0.2  \\
$SCM_{5}$ &  0.0 & 0.4 & 0.52 & 0.55 & 0.55 & 0.0 &  0.55 & $\Leftrightarrow$ & 0.0 \\
$SCM_{6}$ &  -0.2 & 0.6 & 0.72 & 0.55 & 0.71  & -0.2 &  0.71 & $\Leftrightarrow$ & -0.2 \\
    \bottomrule
  \end{tabular}
  \caption{
  Six observationally and/or interventionally dissimilar SCMs with different mixtures of total observed association ($\rho_{P_{Obs}}$), non-causal association ($\rho_{true}$), and causal association ($ACE_{true}$). 
  }
  \label{tbl: Jose toy SCM}
\end{table*}
\begin{figure*}[ht!]
\floatconts
  {fig:Jose_HVR_SCMs_050000_rho_curves}
  {\caption{Observationally equivalent SCMs, e.g., $\{SCM_{1},SCM_{2},SCM_{3}\}$ and $\{SCM_{4},SCM_{5},SCM_{6}\}$, show similar scatter plots, and $\rho_{P_{Obs}}$, as well as similar $\rho_{curve}$, confirming respective observational equivalences.}}
  {%
    \subfigure[Scatter plots of the six observational datasets from the respective SCMs.]{\label{sfig:Jose_HVR_SCMs_050000_rho_curves_b}%
      \includegraphics[width=0.37\linewidth]{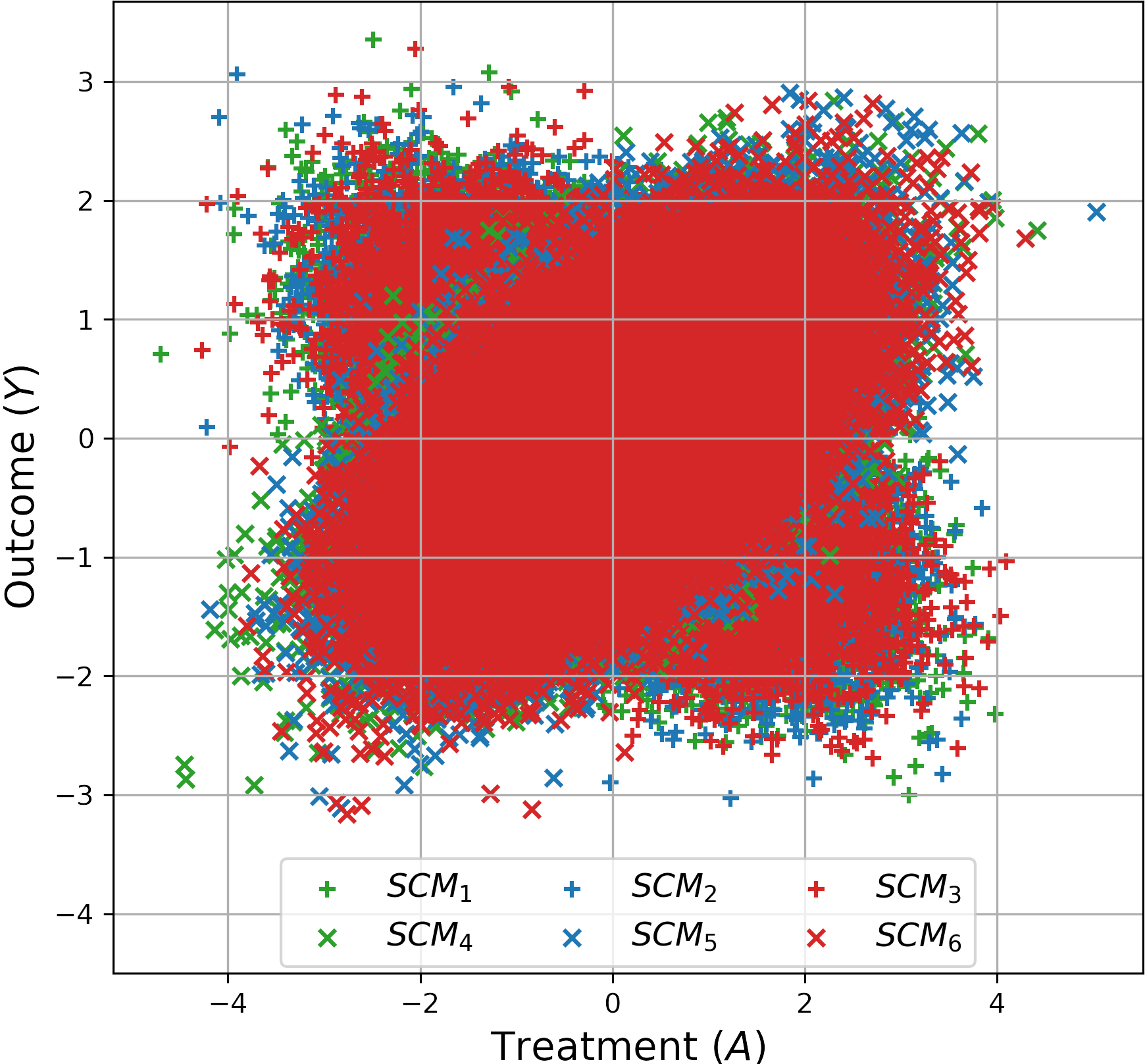}}
    \qquad
    \subfigure[Six $\rho_{curve}$ obtained using observational dataset from six SCMs in Table~\ref{tbl: Jose toy SCM}.]{\label{sfig:Jose_HVR_SCMs_050000_rho_curves_a}%
      \includegraphics[width=0.55\linewidth]{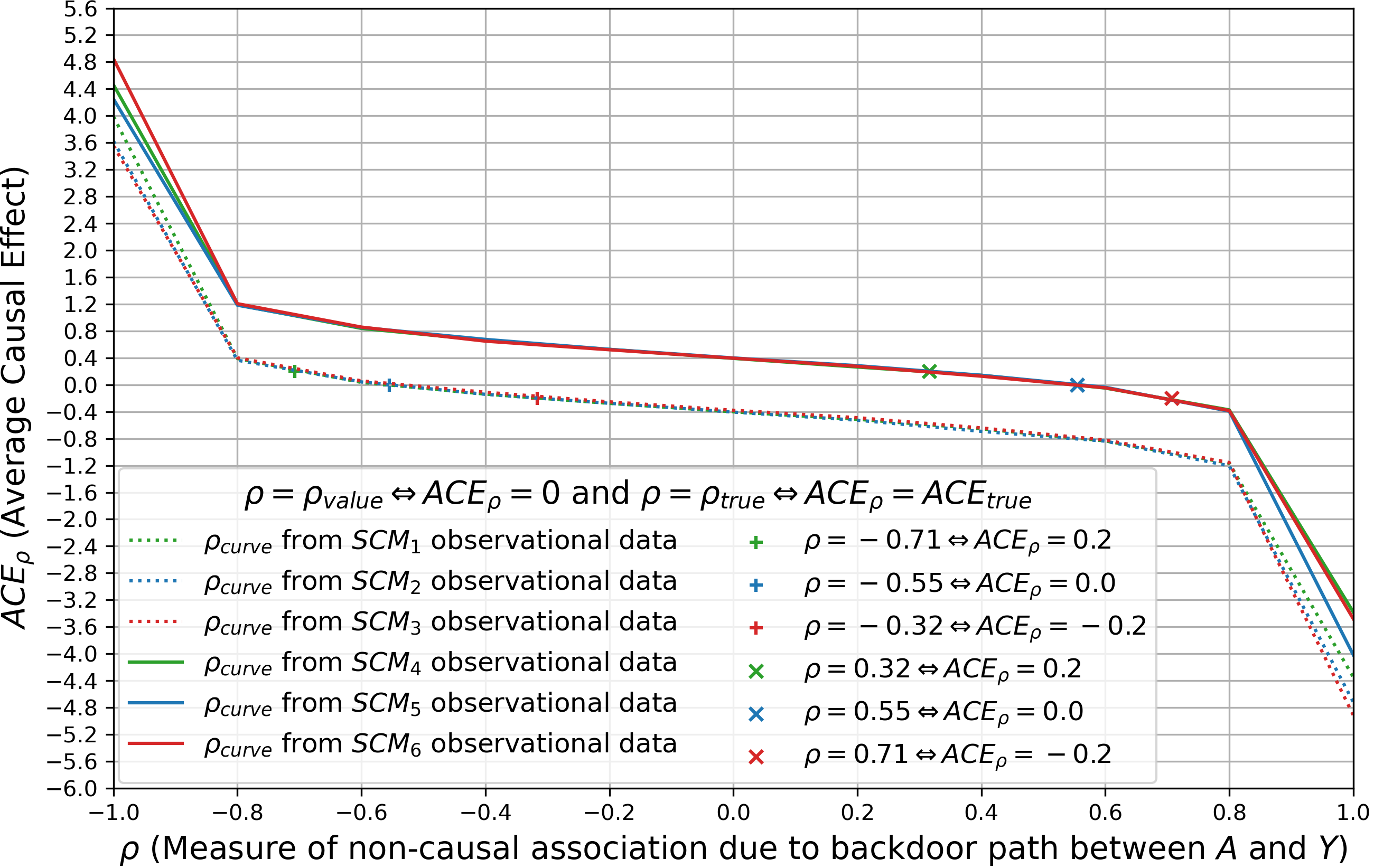}
      }%
  }
\end{figure*}
In our first set of simulated experiments, we consider the SCM with continuous treatment $A$ and outcome $Y$ variables by~\citet{hoover2006causality}. This is a well-studied SCM from economics and econometrics, and is defined as below. The true causal, non-causal, and total observed (causal+non-causal) associations in the SCM are known and are used for verification purposes.

\vspace{-5mm}
\begin{IEEEeqnarray}{C}
SCM_{\alpha,\beta,\delta} \enspace:: \enspace A {:=}\varepsilon_A   \enspace, \enspace\enspace Y{:=}\alpha A{+}\varepsilon_Y \enspace , \begin{bmatrix}
\varepsilon_A\\
\varepsilon_Y
\end{bmatrix} \sim \mathcal{N} \left(\begin{bmatrix}
0 \\
0
\end{bmatrix},\begin{bmatrix}
1 & \beta \\
\beta & \delta
\end{bmatrix}\right) \enspace.\IEEEyesnumber\label{seq:Hoover Jose toy SCM}
\end{IEEEeqnarray}
For a given set of SCM parameters $\alpha,\beta,\delta$ in Eq.~\eqref{seq:Hoover Jose toy SCM}, we have the corresponding Pearson's correlation $\textcolor{darkgreen}{\rho_{P_{Obs}}}{=}\frac{\sigma_{A,Y}}{\sigma_{A}\sigma_{Y}}$, $\sigma^2_A{=}1$, $\sigma^2_Y{=}\alpha^2{+}\delta{+}2\alpha\beta$, $\sigma_{A,Y}{=}\alpha{+}\beta$, $\rho_{true}{=}\beta/\delta$ and $ACE_{true}{=}\alpha$ as tabulated in Table~\ref{tbl: Jose toy SCM}. 
Figure~\ref{sfig:Jose_HVR_SCMs_050000_rho_curves_b} indicates the scatter plot of each of the six observational datasets from the SCM in Eq.~\eqref{seq:Hoover Jose toy SCM} as parameterized in Table~\ref{tbl: Jose toy SCM}. 
The SCMs that are observationally equivalent with the same $\textcolor{darkgreen}{\rho_{P_{Obs}}}$ present similar observational data distributions even though these data distributions have been generated from interventionally dissimilar SCMs.
Figure~\ref{sfig:Jose_HVR_SCMs_050000_rho_curves_a} shows observationally equivalent datasets result in equivalent $\rho_{curve}$, e.g., $\rho_{curve}$ of $\{SCM_1,SCM_2,SCM_3\}$ are similar since they are observationally equivalent as verifiable in the scatter plots of their respective observational dataset from  Figure~\ref{sfig:Jose_HVR_SCMs_050000_rho_curves_b}.
From Figure~\ref{sfig:Jose_HVR_SCMs_050000_rho_curves_a} and Table~\ref{tbl: Jose toy SCM}, we empirically observe the results $\rho{=}\textcolor{darkgreen}{\rho_{value}}$ corresponds to $ACE_{\rho}{=}0$ and $\rho{=}\rho_{true}$ corresponds to $ACE_{\rho}{=}ACE_{true}$ as presented in Section~\ref{sec:evalue_rho_value}. In summary, the main contribution of our method is the estimation of ACE as a function of the sensitivity parameter $\rho$ that indicates the degree of the confounding, thus distinguishing observationally equivalent SCMs based on the assumed degree of unconfoundedness.

The SCM in Eq.~\eqref{seq:Hoover Jose toy SCM} used to illustrate $\rho$-GNF with continuous outcomes may be considered a simple case, i.e. a linear SCM similar to the ones studied in~\citet{cinelli2019sensitivity, cinelli2020making}. However, as the $\rho$-GNF is parameterized to learn arbitrary non-linear monotonic functions using UMNN transformers~\citep{wehenkel2019UMNN}, our work is a generalization of the sensitivity analysis for simple linear SCMs to complex non-linear SCMs, as we demonstrate in the subsequent experiments with binary and categorical outcomes in Section~\ref{sec:experiments simbin}.

\subsection{Experiments with Discrete (Binary/Categorical) Outcomes}\label{sec:experiments simbin}
\begin{figure*}[ht!]
    \centering
    \includegraphics[width=\linewidth,height=\linewidth,keepaspectratio]{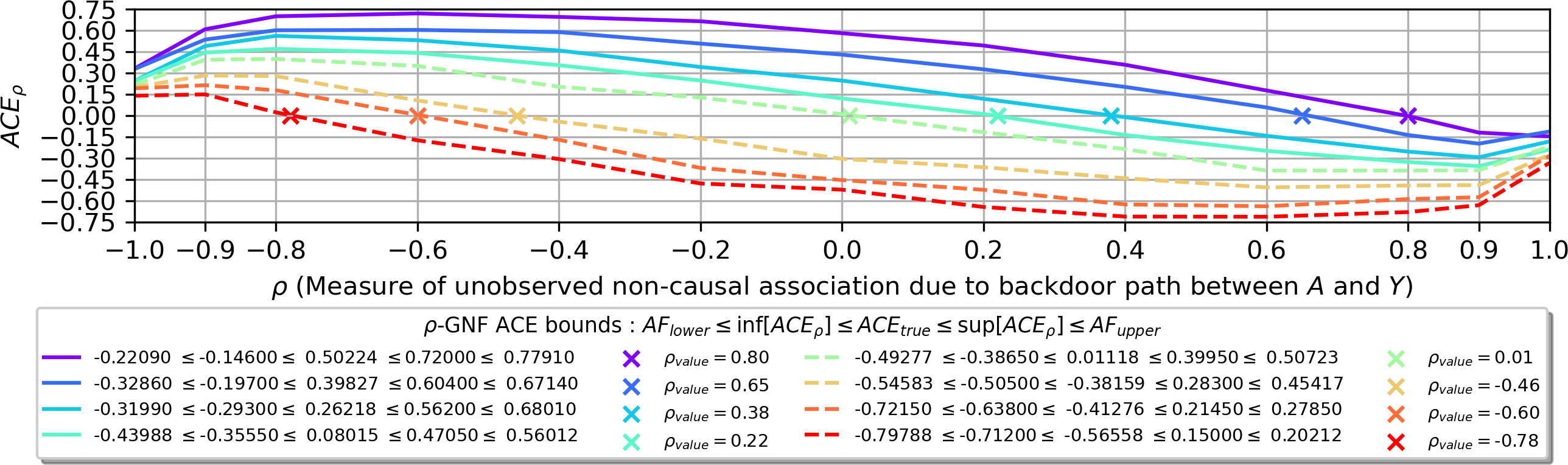}
    \caption{
    Eight $\rho_{curve}$ each corresponding to a randomly generated dataset from a randomly parameterized DGP, and their respective $\rho_{value}$. 
    The true ACE identified using $do$-calculus is indeed included for all the $\rho_{curve}$ with our empirical bounds narrower than the AF bounds.
    }
    \label{fig:ndgp_0000_9500_subset}
\end{figure*}
In the second set of experiments, we consider multiple (twenty) randomly generated observational datasets from randomly sampled data generating processes as proposed in~\citet{sjolander2020note, sjolander2021novel, pena2022simple}, where all variables are binary. As proposed, we randomly sample the parameters $\{\mathbb{P}(U), \mathbb{P}(A|U), \mathbb{P}(Y|A,U)\}$ of the binary data generating process (DGP) from the uniform distribution in [0,1], where $U$ denotes the confounder of $A$ and $Y$ which is intentionally hidden during training to simulate unobserved confounding. 
For this simple binary outcome/treatment/confounder case, the gold-standard AF bounds of a constant width of 1 $(=p_1{+}p_0)$ with 100\% certainty of including the true ACE are presented in~\citet{robins1989analysis, manski1990sensitivitybounds} as

\vspace{-5mm}\begin{IEEEeqnarray}{lC}
\textrm{AF ACE bounds}  : \enspace& AF_{lower}{=}q_1p_1{-}q_0p_0{-}p_1 \leq ACE_{true}\leq AF_{upper}{=}q_1p_1{-}q_0p_0{+}p_0
\enspace, \IEEEyesnumber\label{seq:AFbounds}
\end{IEEEeqnarray}
where $p_a{=}\mathbb{P}(A{=}a)$ and $q_a{=}\mathbb{P}(Y{=}1|A{=}a)$ are estimated using observational data.
The $\rho_{curve}$ in Figure~\ref{fig:ndgp_0000_9500_subset} does include the true ACE in all of the randomly generated DGPs. 
Further, we obtain a narrower bound of 0.83$\pm$0.03 (i.e., $\approx$14\% narrower than the AF bounds) such that our empirical bounds lie within the AF bounds, unlike the bounds by~\citet{vanderweele2017sensitivity} which may be wider than the AF bounds, as shown in~\citet{sjolander2020note}. 
The former comes as no surprise: Since our bounds are empirically obtained particularly assuming the Gaussian copula, they must lie within the AF bounds. 
Figure~\ref{fig:ndgp_0000_9500_subset} verifies it, i.e.,

\vspace{-5mm}\begin{IEEEeqnarray}{lC}
\rho\textrm{-GNF ACE bounds}  : \enspace & AF_{lower} \leq \inf[ACE_{\rho}] \leq ACE_{true}\leq \sup[ACE_{\rho}] \leq AF_{upper}\enspace. \IEEEyesnumber\label{seq:total_bounds}
\end{IEEEeqnarray}

\begin{figure*}[ht!]
    \centering
    \includegraphics[width=\linewidth,height=\linewidth,keepaspectratio]{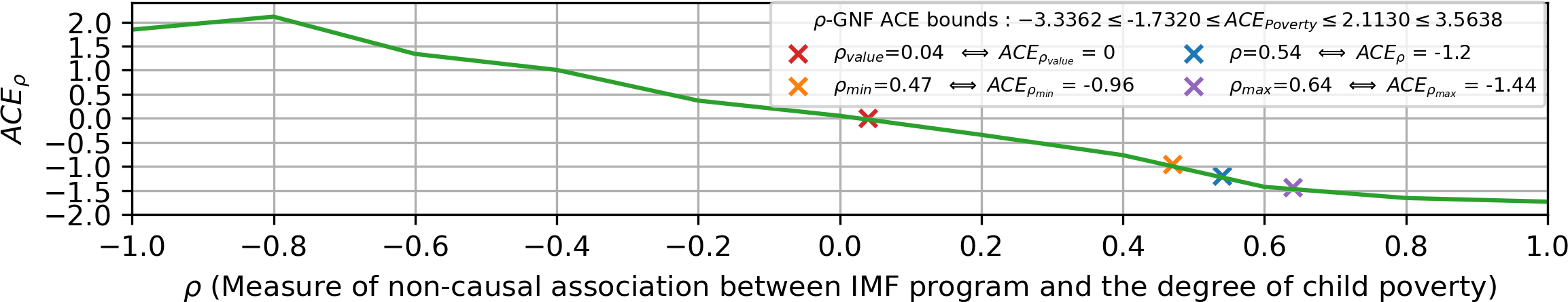}
    \caption{
    $\rho_{curve}$ for the total degree of the child poverty along with the bounds from Eq.~\eqref{seq:total_bounds}.
    The point $(\rho,ACE_{\rho}){=}(0.54, -1.2)$ corresponds to the ACE estimate from~\citet{balgi2022counterfactual} obtained with observed confounders and unconfoundedness as well as the ACE interval $-1.2\pm0.24$ respectively corresponds to $[+0.47, +0.64]$ interval of $\rho$.
    }
    \label{fig:IMFCP_sensitivity_analysis_full}
\end{figure*}
Subsequently, we also extend the experiment from binary outcome to categorical outcome using a real-world dataset with categorical outcomes analysing the IMF (International Monetary Fund) program impact on the degree of child poverty in the Global-South region with the outcome as the total degree of child poverty ranging in 0-7 classes~\citep{balgi2022counterfactual}.
Unlike the binary outcomes, the AF bounds are not available for the non-binary outcomes such as the categorical total degree of child poverty (degree 0: no poverty to degree 7: severe poverty).
Since the degree of child poverty is formulated as the sum of seven binary individual dimension of child poverty, we may identify the AF bounds for each binary individual dimension of child poverty and extend the AF bounds to the total degree of child poverty.
From Eq.~\eqref{seq:AFbounds} and the observational dataset of the seven binary individual dimensions of child poverty, the assumption-free lower and upper bounds are identified as, 
\begin{inparaenum}[(i)]
\item education: $[-0.4843, 0.5157]$,
\item health: $[-0.5009, 0.4991]$, 
\item information: $[-0.4893, 0.5107]$,
\item malnutrition: $[-0.5111, 0.4889]$,
\item sanitization: $[-0.5012, 0.4988]$,
\item shelter: $[-0.4756, 0.5244]$,
\item water: $[-0.4738, 0.5262]$.
\end{inparaenum}

Since, the total degree of child poverty is the sum of these seven binary individual dimensions of child poverty, the lower and upper assumption-free bounds total degree of child poverty can be identified as sum of the respective lower and upper bounds. 
Thus, obtain the AF bounds for the total degree of child poverty as $[-3.3362,+3.5638]$ as indicated in the $\rho_{curve}$ in Figure~\ref{fig:IMFCP_sensitivity_analysis_full}. Note that since the binary AF bounds are of width 1, and the total degree of child poverty is a sum of seven binary variables, the AF bounds for the total degree of the child poverty is of width 7.
From Figure~\ref{fig:IMFCP_sensitivity_analysis_full}, our ACE bounds $[-1.73, +2.11]$ are verified empirically with a narrower width of 3.84, i.e., a reduction of the bounds by $45.1\%$ under the Gaussian copula assumption.

\section{Conclusion}\label{sec:conslusion}
\looseness=-1
We proposed a novel copula-based approach for sensitivity analysis termed $\rho$-GNF to bound the causal effect in the risk difference scale where $\rho{\in}[-1,+1]$ is a bounded sensitivity parameter representing the unobserved back-door non-causal association between the observed treatment and outcome. 
Under the Gaussian copula assumption, we showed that $\rho$-GNF enabled us to estimate the causal effect as a function of $\rho$ in the form of $\rho_{curve}$. 
The $\rho_{curve}$ enabled us to identify narrower empirical bounds in contrast to the wider AF bounds, irrespective of discrete or continuous outcome variables. 
We identified $\rho_{value}$ as the measure of the unobserved confounding that explains away the causal effect and presented the similarities to the E-value with experimental validation. 
Further, the $\rho_{curve}$ enabled us to provide finer bounds for the causal effect given an interval of $\rho$ values that the domain expert considers appropriate to identify the sign of the ACE, thus deducing if the treatment is beneficial or harmful. 
The adaptability of $\rho$-GNF for both discrete and continuous outcomes should encourage the use of sensitivity analysis when working with non-randomized observational data to draw causal conclusions.


\bibliography{pgm2024_short}

\begin{thebibliography}{82}
\providecommand{\natexlab}[1]{#1}
\providecommand{\url}[1]{\texttt{#1}}
\expandafter\ifx\csname urlstyle\endcsname\relax
  \providecommand{\doi}[1]{doi: #1}\else
  \providecommand{\doi}{doi: \begingroup \urlstyle{rm}\Url}\fi

\bibitem[Angus(1994)]{angus1994probability}
J.~E. Angus.
\newblock \href{https://doi.org/10.1137/1036146}{The Probability Integral Transform and Related Results}.
\newblock \emph{SIAM review}, 36\penalty0 (4):\penalty0 652--654, 1994.

\bibitem[Balgi et~al.(2022{\natexlab{a}})Balgi, Pe{\~n}a, and Daoud]{balgi2022cgnf}
S.~Balgi, J.~M. Pe{\~n}a, and A.~Daoud.
\newblock \href{https://ojs.aaai.org/index.php/AAAI/article/view/21437}{Personalized Public Policy Analysis in Social Sciences Using Causal-Graphical Normalizing Flows}.
\newblock In \emph{AAAI Conference on Artificial Intelligence (AAAI)}, pages 11810--11818, 2022{\natexlab{a}}.

\bibitem[Balgi et~al.(2022{\natexlab{b}})Balgi, Pe{\~n}a, and Daoud]{balgi2022counterfactual}
S.~Balgi, J.~M. Pe{\~n}a, and A.~Daoud.
\newblock \href{https://doi.org/10.48550/arXiv.2202.09391}{Counterfactual Analysis of the Impact of the {IMF} Program on Child Poverty in the Global-South Region using Causal-Graphical Normalizing Flows}.
\newblock \emph{arXiv preprint arXiv:2202.09391}, 2022{\natexlab{b}}.

\bibitem[Benali et~al.(2021)Benali, Bod{\'e}n{\`e}s, Labroche, and de~Runz]{benali2021mtcopula}
F.~Benali, D.~Bod{\'e}n{\`e}s, N.~Labroche, and C.~de~Runz.
\newblock \href{https://hal.archives-ouvertes.fr/hal-03188317}{{MTC}opula: Synthetic Complex Data Generation Using Copula}.
\newblock In \emph{International Workshop on Design, Optimization, Languages and Analytical Processing of Big Data (DOLAP)}, pages 51--60, 2021.

\bibitem[Brumback et~al.(2004)Brumback, Hern{\'a}n, Haneuse, and Robins]{brumback2004sensitivity}
B.~A. Brumback, M.~A. Hern{\'a}n, S.~J. Haneuse, and J.~M. Robins.
\newblock \href{https://doi.org/10.1002/sim.1657}{Sensitivity Analyses for Unmeasured Confounding Assuming a Marginal Structural Model for Repeated Measures}.
\newblock \emph{Statistics in Medicine}, 23\penalty0 (5):\penalty0 749--767, 2004.

\bibitem[Cherubini et~al.(2004)Cherubini, Luciano, and Vecchiato]{cherubini2004copula}
U.~Cherubini, E.~Luciano, and W.~Vecchiato.
\newblock \emph{\href{https://doi.org/10.1002/9781118673331}{Copula Methods in Finance}}.
\newblock John Wiley \& Sons, 2004.

\bibitem[Cinelli and Hazlett(2020)]{cinelli2020making}
C.~Cinelli and C.~Hazlett.
\newblock \href{http://dx.doi.org/10.1111/rssb.12348}{ Making Sense of Sensitivity: Extending Omitted Variable Bias}.
\newblock \emph{Journal of the Royal Statistical Society: Series B (Statistical Methodology)}, 82\penalty0 (1):\penalty0 39--67, 2020.

\bibitem[Cinelli et~al.(2019)Cinelli, Kumor, Chen, Pearl, and Bareinboim]{cinelli2019sensitivity}
C.~Cinelli, D.~Kumor, B.~Chen, J.~Pearl, and E.~Bareinboim.
\newblock \href{http://proceedings.mlr.press/v97/cinelli19a/cinelli19a.pdf}{Sensitivity Analysis of Linear Structural Causal Models}.
\newblock In \emph{International Conference on Machine Learning (ICML)}, pages 1252--1261, 2019.

\bibitem[Cochran and Rubin(1973)]{cochran1973controlling}
W.~G. Cochran and D.~B. Rubin.
\newblock \href{https://www.jstor.org/stable/25049893}{Controlling Bias in Observational Studies: A Review}.
\newblock \emph{Sankhy{\=a}: The Indian Journal of Statistics, Series A}, pages 417--446, 1973.

\bibitem[Cornfield et~al.(1959)Cornfield, Haenszel, Hammond, Lilienfeld, Shimkin, and Wynder]{cornfield1959smoking}
J.~Cornfield, W.~Haenszel, E.~C. Hammond, A.~M. Lilienfeld, M.~B. Shimkin, and E.~L. Wynder.
\newblock \href{https://doi.org/10.1093/ije/dyp289}{Smoking and Lung Cancer: Recent Evidence and a Discussion of Some Questions}.
\newblock \emph{Journal of the National Cancer Institute (JNCI)}, 22\penalty0 (1):\penalty0 173--203, 1959.

\bibitem[Cox(1958)]{cox1958planning}
D.~R. Cox.
\newblock \emph{\href{https://doi.org/10.1017/S0020268100038063}{Planning of Experiments}}.
\newblock New York: Wiley, 1958.

\bibitem[Ding and VanderWeele(2016)]{ding2016sensitivity}
P.~Ding and T.~J. VanderWeele.
\newblock \href{https://doi.org/10.1097/ede.0000000000000457}{Sensitivity Analysis Without Assumptions}.
\newblock \emph{Epidemiology (Cambridge, Mass.)}, 27\penalty0 (3):\penalty0 368, 2016.

\bibitem[D'Onofrio et~al.(2020)D'Onofrio, Sj{\"o}lander, Lahey, Lichtenstein, and {\"O}berg]{d2020accounting}
B.~M. D'Onofrio, A.~Sj{\"o}lander, B.~B. Lahey, P.~Lichtenstein, and A.~S. {\"O}berg.
\newblock \href{https://doi.org/10.1146/annurev-clinpsy-032816-045030}{Accounting for confounding in observational studies}.
\newblock \emph{Annual Review of Clinical Psychology}, 16:\penalty0 25--48, 2020.

\bibitem[Durante and Sempi(2016)]{durante2016principles}
F.~Durante and C.~Sempi.
\newblock \emph{\href{https://doi.org/10.1201/b18674}{Principles of Copula Theory}}, volume 474.
\newblock CRC press Boca Raton, 2016.

\bibitem[Fienberg and Duncan(1975)]{Fienberg1975Introduction2SEM}
S.~E. Fienberg and O.~D. Duncan.
\newblock \href{https://doi.org/10.2307/2286831}{Introduction to Structural Equation Models}.
\newblock \emph{Journal of the American Statistical Association (JASA)}, 72:\penalty0 485, 1975.

\bibitem[Fisher(1936)]{fisher1936designofexperiments}
R.~A. Fisher.
\newblock \href{https://doi.org/10.1136/bmj.1.3923.554-a}{Design of Experiments}.
\newblock \emph{British Medical Journal (BMJ)}, 1:\penalty0 554--554, 1936.

\bibitem[Goldberger(1972)]{Goldberger1972SEMecon}
A.~S. Goldberger.
\newblock \href{https://doi.org/10.2307/1913851}{Structural Equation Methods in the Social Sciences}.
\newblock \emph{Econometrica}, 40:\penalty0 979--1001, 1972.

\bibitem[Haavelmo(1943)]{Haavelmo1943SEM}
T.~Haavelmo.
\newblock \href{https://doi.org/10.2307/1905714}{The Statistical Implications of a System of Simultaneous Equations}.
\newblock \emph{Econometrica}, 11:\penalty0 1--12, 1943.

\bibitem[Hern{\'a}n and Robins(2009)]{hernan2009ipw}
M.~A. Hern{\'a}n and J.~M. Robins.
\newblock \emph{\href{https://cdn1.sph.harvard.edu/wp-content/uploads/sites/1268/2022/10/hernanrobins_WhatIf_15sep22.pdf}{Causal Inference: What If}}.
\newblock Boca Raton: Chapman \& Hall/CRC, 2009.

\bibitem[Hoover(2006)]{hoover2006causality}
K.~D. Hoover.
\newblock \href{https://dx.doi.org/10.2139/ssrn.930739}{Causality in Economics and Econometrics}.
\newblock \emph{SSRN eLibrary}, 2006.

\bibitem[Huang et~al.(2018)Huang, Krueger, Lacoste, and Courville]{Huang2018NAF}
C.~Huang, D.~Krueger, A.~Lacoste, and A.~C. Courville.
\newblock \href{http://proceedings.mlr.press/v80/huang18d.html}{Neural Autoregressive Flows}.
\newblock In \emph{International Conference on Machine Learning (ICML)}, pages 2083--2092, 2018.

\bibitem[Huang and Valtorta(2006)]{HuangV06docalculus}
Y.~Huang and M.~Valtorta.
\newblock \href{https://dl.acm.org/doi/10.5555/3020419.3020446}{Pearl's Calculus of Intervention Is Complete}.
\newblock In \emph{Uncertainty in Artificial Intelligence (UAI)}, 2006.

\bibitem[Ilse et~al.(2021)Ilse, Forr{\'e}, Welling, and Mooij]{ilse2021efficient}
M.~Ilse, P.~Forr{\'e}, M.~Welling, and J.~M. Mooij.
\newblock \href{https://doi.org/10.48550/arXiv.2103.04786}{Combining Interventional and Observational Data Using Causal Reductions}.
\newblock \emph{arXiv preprint arXiv:2103.04786}, 2021.

\bibitem[Imbens(2003)]{imbens2003sensitivity}
G.~W. Imbens.
\newblock \href{https://doi.org/10.1257/000282803321946921}{Sensitivity to Exogeneity Assumptions in Program Evaluation}.
\newblock \emph{American Economic Review}, 93\penalty0 (2):\penalty0 126--132, 2003.

\bibitem[Imbens and Rubin(2015)]{Imbens2015CIinSocialscience}
G.~W. Imbens and D.~B. Rubin.
\newblock \emph{\href{https://doi.org/10.1017/cbo9781139025751}{Causal Inference for Statistics, Social, and Biomedical Sciences: An Introduction}}.
\newblock Cambridge University Press, USA, 2015.

\bibitem[Ioannidis et~al.(2019)Ioannidis, Tan, and Blum]{ioannidis2019limitations}
J.~Ioannidis, Y.~Tan, and M.~Blum.
\newblock \href{https://doi.org/10.7326/M18-2159}{Limitations and Misinterpretations of {E}-values for Sensitivity Analyses of Observational Studies}.
\newblock \emph{Internal Medicine}, 170\penalty0 (2):\penalty0 108--111, 2019.

\bibitem[Javaloy et~al.(2023)Javaloy, S{\'{a}}nchez{-}Mart{\'{\i}}n, and Valera]{JavaloySV23}
A.~Javaloy, P.~S{\'{a}}nchez{-}Mart{\'{\i}}n, and I.~Valera.
\newblock \href{https://doi.org/10.48550/arXiv.2306.05415}{Causal Normalizing Flows: From Theory to Practice}.
\newblock In \emph{Neural Information Processing Systems (NeurIPS)}, 2023.

\bibitem[Kendall(1938)]{kendall1938new}
M.~G. Kendall.
\newblock \href{https://doi.org/10.1093/biomet/30.1-2.81}{A New Measure of Rank Correlation}.
\newblock \emph{Biometrika}, 30\penalty0 (1/2):\penalty0 81--93, 1938.

\bibitem[Kobyzev et~al.(2021)Kobyzev, Prince, and Brubaker]{kobyzev2020NF}
I.~Kobyzev, S.~Prince, and M.~Brubaker.
\newblock \href{https://doi.org/10.1109/TPAMI.2020.2992934}{Normalizing Flows: An Introduction and Review of Current Methods}.
\newblock \emph{IEEE Transactions on Pattern Analysis and Machine Intelligence (TPAMI)}, 43\penalty0 (11):\penalty0 3964--3979, 2021.

\bibitem[Kruskal(1958)]{kruskal1958ordinal}
W.~H. Kruskal.
\newblock \href{https://doi.org/10.2307/2281954}{Ordinal Measures of Association}.
\newblock \emph{Journal of the American Statistical Association (JASA)}, 53\penalty0 (284):\penalty0 814--861, 1958.

\bibitem[Kumar(2019)]{kumar2019copula}
P.~Kumar.
\newblock \href{https://doi.org/10.1007/978-981-13-0872-7}{Copula Functions and Applications in Engineering}.
\newblock In \emph{Logistics, Supply Chain and Financial Predictive Analytics}, pages 195--209. Springer, 2019.

\bibitem[Lash et~al.(2009)Lash, Fox, Fink, et~al.]{lash2009applying}
T.~L. Lash, M.~P. Fox, A.~K. Fink, et~al.
\newblock \emph{\href{https://doi.org/10.1007/978-0-387-87959-8}{Applying Quantitative Bias Analysis to Epidemiologic Data}}.
\newblock Springer, 2009.

\bibitem[Lindmark et~al.(2018)Lindmark, de~Luna, and Eriksson]{lindmark2018sensitivity}
A.~Lindmark, X.~de~Luna, and M.~Eriksson.
\newblock \href{https://doi.org/10.1002/sim.7620}{Sensitivity Analysis for Unobserved Confounding of Direct and Indirect Effects Using Uncertainty Intervals}.
\newblock \emph{Statistics in Medicine}, 37\penalty0 (10):\penalty0 1744--1762, 2018.

\bibitem[Ling et~al.(2020)Ling, Fang, and Kolter]{ling2020deep}
C.~K. Ling, F.~Fang, and J.~Z. Kolter.
\newblock \href{https://papers.nips.cc/paper/2020/hash/10eb6500bd1e4a3704818012a1593cc3-Abstract.html}{Deep Archimedean Copulas}.
\newblock \emph{Neural Information Processing Systems (NeurIPS)}, 33:\penalty0 1535--1545, 2020.

\bibitem[MacKenzie and Spears(2014)]{mackenzie2014formula}
D.~MacKenzie and T.~Spears.
\newblock \href{https://www.jstor.org/stable/43284238}{The Formula That Killed Wall Street: The Gaussian Copula and Modelling Practices in Investment Banking}.
\newblock \emph{Social Studies of Science}, 44\penalty0 (3):\penalty0 393--417, 2014.

\bibitem[Manski(1990)]{manski1990sensitivitybounds}
C.~F. Manski.
\newblock \href{http://www.jstor.org/stable/2006592}{Nonparametric Bounds on Treatment Effects}.
\newblock \emph{American Economic Review}, 80\penalty0 (2):\penalty0 319--323, 1990.

\bibitem[Mathur et~al.(2018)Mathur, Ding, Riddell, and VanderWeele]{mathur2018website}
M.~B. Mathur, P.~Ding, C.~A. Riddell, and T.~J. VanderWeele.
\newblock \href{https://www.evalue-calculator.com/}{Website and R Package for Computing {E}-values}.
\newblock \emph{Epidemiology}, 29\penalty0 (5):\penalty0 e45, 2018.

\bibitem[Meyer(2013)]{meyer2013bivariate}
C.~Meyer.
\newblock \href{https://doi.org/10.1080/03610926.2011.611316}{The Bivariate Normal Copula}.
\newblock \emph{Communications in Statistics-Theory and Methods}, 42\penalty0 (13):\penalty0 2402--2422, 2013.

\bibitem[Mooij et~al.(2016)Mooij, Peters, Janzing, Zscheischler, and Sch{\"o}lkopf]{mooij2016distinguishing}
J.~M. Mooij, J.~Peters, D.~Janzing, J.~Zscheischler, and B.~Sch{\"o}lkopf.
\newblock \href{https://dl.acm.org/doi/10.5555/2946645.2946677}{Distinguishing Cause From Effect Using Observational Data: Methods and Benchmarks}.
\newblock \emph{Journal of Machine Learning Research (JMLR)}, 17\penalty0 (1):\penalty0 1103--1204, 2016.

\bibitem[Nelsen(2007)]{nelsen2007introduction}
R.~B. Nelsen.
\newblock \emph{\href{https://doi.org/10.1007/0-387-28678-0}{An Introduction to Copulas}}.
\newblock Springer Science \& Business Media, 2007.

\bibitem[Papamakarios et~al.(2017)Papamakarios, Murray, and Pavlakou]{Papamakarios2017MAF}
G.~Papamakarios, I.~Murray, and T.~Pavlakou.
\newblock \href{https://dl.acm.org/doi/10.5555/3294771.3294994}{Masked Autoregressive Flow for Density Estimation}.
\newblock In \emph{Neural Information Processing Systems (NeurIPS)}, pages 2338--2347, 2017.

\bibitem[Papamakarios et~al.(2021)Papamakarios, Nalisnick, Rezende, Mohamed, and Lakshminarayanan]{papamakarios2021NF_pmi}
G.~Papamakarios, E.~Nalisnick, D.~J. Rezende, S.~Mohamed, and B.~Lakshminarayanan.
\newblock \href{https://dl.acm.org/doi/abs/10.5555/3546258.3546315}{Normalizing Flows for Probabilistic Modeling and Inference}.
\newblock \emph{Journal of Machine Learning Research (JMLR)}, 22\penalty0 (57):\penalty0 1--64, 2021.

\bibitem[Paszke et~al.(2017)Paszke, Gross, Chintala, Chanan, Yang, DeVito, Lin, Desmaison, Antiga, and Lerer]{paszke2017pytorch}
A.~Paszke, S.~Gross, S.~Chintala, G.~Chanan, E.~Yang, Z.~DeVito, Z.~Lin, A.~Desmaison, L.~Antiga, and A.~Lerer.
\newblock {\href{https://openreview.net/forum?id=BJJsrmfCZ}{Automatic Differentiation in PyTorch}}.
\newblock \emph{NeurIPS Workshops}, 2017.

\bibitem[Pearl(1999)]{pearl1999probabilitiesofcausation}
J.~Pearl.
\newblock \href{https://doi.org/10.1023/A:1005233831499}{Probabilities of Causation: Three Counterfactual Interpretations and Their Identification}.
\newblock \emph{Synthese}, 121\penalty0 (1):\penalty0 93--149, 1999.

\bibitem[Pearl(2009{\natexlab{a}})]{pearl2009abductionactionprediction}
J.~Pearl.
\newblock \href{https://doi.org/10.1214/09-SS057}{Causal Inference in Statistics: An Overview}.
\newblock \emph{Statistics Surveys}, 3:\penalty0 96--146, 2009{\natexlab{a}}.

\bibitem[Pearl(2009{\natexlab{b}})]{pearl2009causality}
J.~Pearl.
\newblock \emph{\href{https://dl.acm.org/doi/book/10.5555/1642718}{Causality: Models, Reasoning and Inference}}.
\newblock Cambridge University Press, USA, 2009{\natexlab{b}}.

\bibitem[Pearl(2012)]{pearl2012docalculus}
J.~Pearl.
\newblock \href{https://dl.acm.org/doi/abs/10.5555/3020652.3020654}{The $Do$-Calculus Revisited}.
\newblock In \emph{Uncertainty in Artificial Intelligence (UAI)}, page 3–11, 2012.

\bibitem[Pearl and Mackenzie(2018)]{pearl2018bookofwhy}
J.~Pearl and D.~Mackenzie.
\newblock \emph{\href{https://dl.acm.org/doi/book/10.5555/3238230}{The Book of Why: The New Science of Cause and Effect}}.
\newblock Basic Books, Inc., 2018.

\bibitem[Pe{\~n}a(2022)]{pena2022simple}
J.~M. Pe{\~n}a.
\newblock \href{https://doi.org/10.1515/jci-2021-0041}{Simple Yet Sharp Sensitivity Analysis for Unmeasured Confounding}.
\newblock \emph{Journal of Causal Inference (JCI)}, 10\penalty0 (1):\penalty0 1--17, 2022.

\bibitem[Peters et~al.(2017)Peters, Janzing, and Sch{\"o}lkopf]{peters2017eci}
J.~Peters, D.~Janzing, and B.~Sch{\"o}lkopf.
\newblock \emph{\href{https://library.oapen.org/bitstream/handle/20.500.12657/26040/11283.pdf}{Elements of Causal Inference: Foundations and Learning Algorithms}}.
\newblock The MIT Press, 2017.

\bibitem[Renard and Lang(2007)]{renard2007use}
B.~Renard and M.~Lang.
\newblock \href{https://doi.org/10.1016/j.advwatres.2006.08.001}{Use of a Gaussian Copula for Multivariate Extreme Value Analysis: Some Case Studies in Hydrology}.
\newblock \emph{Water Resources}, 30\penalty0 (4):\penalty0 897--912, 2007.

\bibitem[Rezende and Mohamed(2015)]{rezende2015variationalNF}
D.~Rezende and S.~Mohamed.
\newblock \href{https://dl.acm.org/doi/10.5555/3045118.3045281}{Variational Inference with Normalizing Flows}.
\newblock In \emph{International Conference on Machine Learning (ICML)}, pages 1530--1538, 2015.

\bibitem[Robins(1986)]{ROBINS1986gcom}
J.~M. Robins.
\newblock \href{https://doi.org/10.1016/0270-0255(86)90088-6}{A New Approach to Causal Inference in Mortality Studies with a Sustained Exposure Period—Application to Control of the Healthy Worker Survivor Effect}.
\newblock \emph{Mathematical Modelling}, 7\penalty0 (9):\penalty0 1393--1512, 1986.

\bibitem[Robins(1989)]{robins1989analysis}
J.~M. Robins.
\newblock \href{https://cdn1.sph.harvard.edu/wp-content/uploads/sites/343/2013/03/nchsr.pdf}{The Analysis of Randomized and Non-randomized AIDS Treatment Trials Using a New Approach to Causal Inference in Longitudinal Studies}.
\newblock \emph{Health Service Research Methodology: A Focus on AIDS}, pages 113--159, 1989.

\bibitem[Robins and Hern{\'a}n(2008)]{robins2008ipw}
J.~M. Robins and M.~A. Hern{\'a}n.
\newblock \href{https://doi.org/10.1371/journal.pgen.1010290}{Estimation of the Causal Effects of Time-varying Exposure}.
\newblock \emph{Longitudinal Data Analysis (LDA)}, pages 553--599, 2008.

\bibitem[Rothman et~al.(2008)Rothman, Greenland, Lash, et~al.]{rothman2008modern}
K.~J. Rothman, S.~Greenland, T.~L. Lash, et~al.
\newblock \emph{Modern Epidemiology}, volume~3.
\newblock Wolters Kluwer Health/Lippincott Williams \& Wilkins Philadelphia, 2008.

\bibitem[Rubin(1990)]{Rubin1990unconfoundedness}
D.~B. Rubin.
\newblock \href{https://doi.org/10.1016/0378-3758(90)90077-8}{Formal Mode of Statistical Inference for Causal Effects}.
\newblock \emph{Journal of Statistical Planning and Inference (JSPI)}, 25\penalty0 (3):\penalty0 279--292, 1990.

\bibitem[Salmon(2009)]{salmon2009recipe}
F.~Salmon.
\newblock \href{https://doi.org/10.1111/j.1740-9713.2012.00538.x}{Recipe for Disaster: The Formula That Killed Wall Street}.
\newblock \emph{Wired Magazine}, 17\penalty0 (3):\penalty0 17--03, 2009.

\bibitem[Salvadori et~al.(2007)Salvadori, De~Michele, Kottegoda, and Rosso]{salvadori2007extremes}
G.~Salvadori, C.~De~Michele, N.~T. Kottegoda, and R.~Rosso.
\newblock \emph{\href{http://dx.doi.org/10.1007/1-4020-4415-1}{Extremes in Nature: An Approach Using Copulas}}, volume~56.
\newblock Springer Science \& Business Media, 2007.

\bibitem[Schlesselman(1978)]{schlesselman1978assessing}
J.~J. Schlesselman.
\newblock \href{https://doi.org/10.1093/oxfordjournals.aje.a112581}{Assessing Effects of Confounding Variables}.
\newblock \emph{American Journal of Epidemiology}, 108\penalty0 (1):\penalty0 3--8, 1978.

\bibitem[Sj{\"o}lander(2020)]{sjolander2020note}
A.~Sj{\"o}lander.
\newblock \href{https://doi.org/10.1515/jci-2020-0012}{A Note on a Sensitivity Analysis for Unmeasured Confounding, and the Related {E}-value}.
\newblock \emph{Journal of Causal Inference (JCI)}, 8\penalty0 (1):\penalty0 229--248, 2020.

\bibitem[Sj{\"o}lander and Greenland(2022)]{sjolander2022values}
A.~Sj{\"o}lander and S.~Greenland.
\newblock {\href{https://doi.org/10.1093/ije/dyac018}{Are {E}-values Too Optimistic or Too Pessimistic? Both and Neither!}}
\newblock \emph{International Journal of Epidemiology}, 2022.

\bibitem[Sj{\"o}lander and H{\"o}ssjer(2021)]{sjolander2021novel}
A.~Sj{\"o}lander and O.~H{\"o}ssjer.
\newblock \href{https://doi.org/10.1515/jci-2021-0024}{Novel Bounds for Causal Effects Based on Sensitivity Parameters on the Risk Difference Scale}.
\newblock \emph{Journal of Causal Inference (JCI)}, 9\penalty0 (1):\penalty0 190--210, 2021.

\bibitem[Sklar(1959)]{sklar1959fonctions}
A.~Sklar.
\newblock \href{http://www.sciepub.com/reference/94917}{Fonctions De R{\'e}partition {\`a} N Dimensions Et Leurs Marges}.
\newblock \emph{Publ. inst. statist. univ. Paris}, 8:\penalty0 229--231, 1959.

\bibitem[Sklar(1973)]{sklar1973random}
A.~Sklar.
\newblock \href{https://www.kybernetika.cz/content/1973/6/449}{Random Variables, Joint Distribution Functions, and Copulas}.
\newblock \emph{Kybernetika}, 9\penalty0 (6):\penalty0 449--460, 1973.

\bibitem[Spearman(1987)]{spearman1987proof}
C.~Spearman.
\newblock \href{https://psycnet.apa.org/doi/10.2307/1412159}{The Proof and Measurement of Association Between Two Things}.
\newblock \emph{The American Journal of Psychology}, 100\penalty0 (3/4):\penalty0 441--471, 1987.

\bibitem[Spearman(2010)]{spearman2010proof}
C.~Spearman.
\newblock \href{https://doi.org/10.1093/ije/dyq191}{The Proof and Measurement of Association Between Two Things}.
\newblock \emph{International Journal of Epidemiology}, 39\penalty0 (5):\penalty0 1137--1150, 2010.

\bibitem[Tabak and Vanden-Eijnden(2010)]{tabak2010nf}
E.~Tabak and E.~Vanden-Eijnden.
\newblock \href{http://dx.doi.org/10.4310/CMS.2010.v8.n1.a11}{Density Estimation by Dual Ascent of the Log-Likelihood}.
\newblock \emph{Communications in Mathematical Sciences}, 8:\penalty0 217--233, 2010.

\bibitem[Tabak and Turner(2013)]{tabak2013nf}
E.~G. Tabak and C.~V. Turner.
\newblock \href{https://doi.org/10.1002/cpa.21423}{A Family of Nonparametric Density Estimation Algorithms}.
\newblock \emph{Communications on Pure and Applied Mathematics}, 66:\penalty0 145--164, 2013.

\bibitem[Tabar et~al.(2022)Tabar, Jung, Yadav, Chavez, Flores, and Lee]{TabarJYCFL2022ML4PP}
M.~Tabar, W.~Jung, A.~Yadav, O.~W. Chavez, A.~Flores, and D.~Lee.
\newblock \href{https://doi.org/10.24963/ijcai.2022/719}{Forecasting the Number of Tenants At-Risk of Formal Eviction: {A} Machine Learning Approach to Inform Public Policy}.
\newblock In \emph{International Joint Conference on Artificial Intelligence {(IJCAI)}}, pages 5178--5184, 2022.

\bibitem[Takeuchi(2010)]{takeuchi2010constructing}
T.~T. Takeuchi.
\newblock \href{http://dx.doi.org/10.1111/j.1365-2966.2010.16778.x}{Constructing a Bivariate Distribution Function With Given Marginals and Correlation: Application to the Galaxy Luminosity Function}.
\newblock \emph{Monthly Notices of the Royal Astronomical Society}, 406\penalty0 (3):\penalty0 1830--1840, 2010.

\bibitem[Tchetgen and Shpitser(2012)]{tchetgen2012semiparametric}
E.~J.~T. Tchetgen and I.~Shpitser.
\newblock \href{https://doi.org/10.1214/12-AOS990}{Semiparametric Theory for Causal Mediation Analysis: Efficiency Bounds, Multiple Robustness, and Sensitivity Analysis}.
\newblock \emph{Annals of Statistics}, 40\penalty0 (3):\penalty0 1816, 2012.

\bibitem[Tian and Pearl(2002)]{tian2002unconfoundedchildrencriteria}
J.~Tian and J.~Pearl.
\newblock \href{https://ftp.cs.ucla.edu/pub/stat_ser/R290-A.pdf}{A General Identification Condition for Causal Effects}.
\newblock In \emph{AAAI Conference on Artificial Intelligence (AAAI)}, page 567–573, 2002.

\bibitem[VanderWeele and Arah(2011)]{vanderweele2011bias}
T.~J. VanderWeele and O.~A. Arah.
\newblock \href{https://doi.org/10.1097/ede.0b013e3181f74493}{Bias Formulas for Sensitivity Analysis of Unmeasured Confounding for General Outcomes, Treatments, and Confounders}.
\newblock \emph{Epidemiology}, pages 42--52, 2011.

\bibitem[VanderWeele and Ding(2017)]{vanderweele2017sensitivity}
T.~J. VanderWeele and P.~Ding.
\newblock \href{https://doi.org/10.7326/M16-2607}{Sensitivity Analysis in Observational Research: Introducing the {E}-value}.
\newblock \emph{Internal Medicine}, 167\penalty0 (4):\penalty0 268--274, 2017.

\bibitem[Veitch and Zaveri(2020)]{veitch2020sense}
V.~Veitch and A.~Zaveri.
\newblock \href{https://proceedings.neurips.cc/paper/2020/hash/7d265aa7147bd3913fb84c7963a209d1-Abstract.html}{Sense and Sensitivity Analysis: Simple Post-hoc Analysis of Bias Due to Unobserved Confounding}.
\newblock \emph{Neural Information Processing Systems (NeurIPS)}, 33:\penalty0 10999--11009, 2020.

\bibitem[Wehenkel and Louppe(2019)]{wehenkel2019UMNN}
A.~Wehenkel and G.~Louppe.
\newblock \href{https://papers.nips.cc/paper/2019/hash/2a084e55c87b1ebcdaad1f62fdbbac8e-Abstract.html}{Unconstrained Monotonic Neural Networks}.
\newblock In \emph{Neural Information Processing Systems (NeurIPS)}, pages 1545--1555, 2019.

\bibitem[Wehenkel and Louppe(2021)]{wehenkel2020GNF}
A.~Wehenkel and G.~Louppe.
\newblock \href{http://proceedings.mlr.press/v130/wehenkel21a/wehenkel21a.pdf}{Graphical Normalizing Flows}.
\newblock In \emph{International Conference on Artificial Intelligence and Statistics (AISTATS)}, pages 37--45, 2021.

\bibitem[Wright(1921)]{wright1921correlationcausation}
S.~Wright.
\newblock \href{https://naldc.nal.usda.gov/download/IND43966364/pdf}{Correlation and Causation}.
\newblock \emph{Journal of Agricultural Research (JAR)}, 20:\penalty0 557--585, 1921.

\bibitem[Zhang and Singh(2019)]{zhang2019copulas}
L.~Zhang and V.~P. Singh.
\newblock \emph{\href{http://dx.doi.org/10.1017/9781108565103.019}{Copulas and Their Applications in Water Resources Engineering}}.
\newblock Cambridge University Press, 2019.

\bibitem[Zheng et~al.(2021)Zheng, D'Amour, and Franks]{zheng2021copula}
J.~Zheng, A.~D'Amour, and A.~Franks.
\newblock \href{https://www.afranks.com/publication/2021-zheng-copula/}{Copula-based Sensitivity Analysis for Multi-Treatment Causal Inference with Unobserved Confounding}.
\newblock \emph{arXiv preprint arXiv:2102.09412}, 2021.

\bibitem[Zheng et~al.(2022)Zheng, Wu, D'Amour, and Franks]{zheng2022sensitivity}
J.~Zheng, J.~Wu, A.~D'Amour, and A.~Franks.
\newblock \href{https://doi.org/10.48550/arXiv.2208.06552}{Sensitivity to Unobserved Confounding in Studies with Factor-structured Outcomes}.
\newblock \emph{arXiv preprint arXiv:2208.06552}, 2022.

\end{thebibliography}






\end{document}